\documentclass[12pt,preprint]{aastex}
\usepackage{lscape}
\usepackage{rotating}

\newcommand{\bdv}[1]{\mbox{\boldmath$#1$}}

\def\rel{{\rm rel}}
\def\e{{\rm E}}
\def\mas{{\rm mas}}
\def\masyr{\mas\,{\rm yr}^{-1}}
\def\au{{\rm AU}} 
\def\muas{{\mu\rm as}}

\def\kpc{{\rm kpc}}

\def\rel{{\rm rel}}

\def\e{{\rm E}}
\def\bpi{{\bdv{\pi}}}

\def\p{{\partial}}
\def\simge{\mathrel{
   \rlap{\raise 0.511ex \hbox{$>$}}{\lower 0.511ex \hbox{$\sim$}}}}
\def\simle{\mathrel{
   \rlap{\raise 0.511ex \hbox{$<$}}{\lower 0.511ex \hbox{$\sim$}}}}

\begin{document}
\title{OGLE-2005-BLG-071Lb, the Most Massive M-Dwarf Planetary Companion?} 

\author{Subo Dong\altaffilmark{1,2}, Andrew Gould\altaffilmark{1,2}, Andrzej Udalski\altaffilmark{3,4}, Jay Anderson\altaffilmark{5},
G.W. Christie\altaffilmark{1,6}, B.\,S.~Gaudi\altaffilmark{1,2},
\\ and \\
M. Jaroszy{\'n}ski\altaffilmark{4}, M. Kubiak\altaffilmark{4}, M.\,K. Szyma{\'n}ski\altaffilmark{4}, G. Pietrzy{\'n}ski\altaffilmark{4,7}, I. Soszy{\'n}ski\altaffilmark{4}, O. Szewczyk\altaffilmark{7,4}, K. Ulaczyk\altaffilmark{4}, {\L}. Wyrzykowski\altaffilmark{8,4} \\ ({The OGLE Collaboration}), \\
D.\,L.~DePoy\altaffilmark{2}, D.\,B.~Fox\altaffilmark{9}, A. Gal-Yam\altaffilmark{10}, C. Han\altaffilmark{11}, S. L\'epine\altaffilmark{12}, 
J.~McCormick\altaffilmark{13}, E. Ofek\altaffilmark{14}, B.-G. Park\altaffilmark{15}, R.\,W. Pogge\altaffilmark{2}, \\ ({The $\mu$FUN Collaboration}),\\
F. Abe\altaffilmark{16}, D.\,P.~Bennett\altaffilmark{17,18}, I.\,A.~Bond\altaffilmark{19}, T.\,R.~Britton\altaffilmark{20}, A.\,C.~Gilmore\altaffilmark{20}. J.\,B.~Hearnshaw\altaffilmark{20}, 
Y. Itow\altaffilmark{16}, K. Kamiya\altaffilmark{16}, P.\,M. Kilmartin\altaffilmark{21}, A. Korpela\altaffilmark{22}, K. Masuda\altaffilmark{16}, Y. Matsubara\altaffilmark{16}, M. Motomura\altaffilmark{16}, 
Y. Muraki\altaffilmark{23}, S. Nakamura\altaffilmark{16}, 
K. Ohnishi\altaffilmark{24}, C. Okada\altaffilmark{16}, N. Rattenbury\altaffilmark{25}, To. Saito\altaffilmark{26}, T. Sako\altaffilmark{16}, 
M. Sasaki\altaffilmark{16}, D. Sullivan\altaffilmark{22}, 
T. Sumi\altaffilmark{16}, P.\,J. Tristram\altaffilmark{21}, T. Yanagisawa\altaffilmark{27}, P.\,C.\,M. Yock\altaffilmark{28}, T. Yoshoika\altaffilmark{16},\\
({The MOA Collaboration})\\
M.D.~Albrow\altaffilmark{20},  J.P.~Beaulieu\altaffilmark{29},
S.~Brillant\altaffilmark{30}, H. Calitz\altaffilmark{31}, A.~Cassan\altaffilmark{32},  
K. H.~Cook\altaffilmark{33}, Ch.~Coutures\altaffilmark{29},
S.~Dieters\altaffilmark{34}, D.~Dominis Prester\altaffilmark{35}, J.~Donatowicz\altaffilmark{36},
P.~Fouqu\'e\altaffilmark{37}, J.~Greenhill\altaffilmark{34}, K.~Hill\altaffilmark{34},
M.~Hoffman\altaffilmark{31}, K.~Horne\altaffilmark{38}, U.G.~J{\o}rgensen\altaffilmark{39},
S.~Kane\altaffilmark{40}, D.~Kubas\altaffilmark{30}, J.B.~Marquette\altaffilmark{29}, R.~Martin\altaffilmark{41},
P.~Meintjes\altaffilmark{31},  J.~Menzies\altaffilmark{42}, K.R.~Pollard\altaffilmark{20},
K.C.~Sahu\altaffilmark{5},  C.~Vinter\altaffilmark{39}, J.~Wambsganss\altaffilmark{32}, A.~Williams\altaffilmark{41}, M. Bode\altaffilmark{43}, D.M.~Bramich\altaffilmark{44}, 
M. Burgdorf\altaffilmark{43}, C. Snodgrass\altaffilmark{30}, I. Steele\altaffilmark{43},
\\ ({The PLANET/RoboNet Collaborations})\\
Vanessa Doublier\altaffilmark{45},
and
Cedric Foellmi\altaffilmark{46}.
}
\altaffiltext{1}
{Microlensing Follow-Up Network ($\mu$FUN).}
\altaffiltext{2}
{Department of Astronomy, Ohio State University,
140 W.\ 18th Ave., Columbus, OH 43210, USA; dong, gould, gaudi, depoy, 
pogge@astronomy.ohio-state.edu.}
\altaffiltext{3}
{Optical Gravitational Lens Experiment (OGLE).}
\altaffiltext{4}
{Warsaw University Observatory, Al.~Ujazdowskie~4, 00-478~Warszawa,
Poland; udalski, mj, msz, mk, pietrzyn, soszynsk, kulaczyk@astrouw.edu.pl.}
\altaffiltext{5}
{Space Telescope Science Institute, Baltimore MD; jayander@stsci.edu.}
\altaffiltext{6}
{Auckland Observatory, Auckland, New Zealand; gwchristie@christie.org.nz.}
\altaffiltext{7}
{Universidad de Concepci{\'o}n, Departamento de Fisica,
Casilla 160--C, Concepci{\'o}n, Chile; szewczyk@astro-udec.cl.}
\altaffiltext{8}
{Institute of Astronomy, University of Cambridge, Madingley Road,
Cambridge CB3 0HA, UK; wyrzykow@ast.cam.ac.uk.}
\altaffiltext{9}
{Astronomy \& Astrophysics, Pennsylvania State University, 525 Davey Laboratory, University Park, PA 16802; dfox@astro.psu.edu.}
\altaffiltext{10}
{Benoziyo Center for Astrophysics, Weizmann Institute of Science,
76100 Rehovot, Israel; avishay.gal-yam@weizmann.ac.il.}
\altaffiltext{11}{Program of Brain Korea, Department of Physics, 
Chungbuk National University, 410 Seongbong-Rho, Hungduk-Gu, 
Chongju 371-763, Korea; cheongho@astroph.chungbuk.ac.kr.}
\altaffiltext{12}{Department of Astrophysics, Division of Physical Sciences,
American Museum of Natural History, Central Park West at 79th Street,
New York, NY 10024, USA; lepine@amnh.org.}
\altaffiltext{13}{Farm Cove Observatory, Centre for Backyard Astrophysics,
Pakuranga, Auckland New Zealand; farmcoveobs@xtra.co.nz.}
\altaffiltext{14}{Division of Physics, Mathematics and Astronomy, California 
Institute of Technology, Pasadena, CA 91125; eran@astro.caltech.edu.}
\altaffiltext{15}{Korea Astronomy and Space Science Institute, 61-1 
Hwaam-Dong, Yuseong-Gu, Daejeon 305-348, Korea; bgpark@kasi.re.kr.}
\altaffiltext{16}{Solar-Terrestrial Environment Laboratory, Nagoya University, 
Nagoya, 464-8601, Japan.}
\altaffiltext{17}{Department of Physics, 225 Nieuwland Science Hall, Notre Dame 
University, Notre Dame, IN 46556, USA.}
\altaffiltext{18}{Probing Lensing Anomalies NETwork (PLANET) Collaboration.}
\altaffiltext{19}{Institute for Information and Mathematical Sciences, Massey University, 
Private Bag 102-904, Auckland 1330, New Zealand.}
\altaffiltext{20}{University of Canterbury, Department of Physics and Astronomy, 
Private Bag 4800, Christchurch 8020, New Zealand.}
\altaffiltext{21}{Mt. John Observatory, P.O. Box 56, Lake Tekapo 8770, New Zealand.}
\altaffiltext{22}{School of Chemical and Physical Sciences, Victoria University, 
Wellington, New Zealand.}
\altaffiltext{23}{Department of Physics, Konan University, Nishiokamoto 8-9-1, 
Kobe 658-8501, Japan.}
\altaffiltext{24}{Nagano National College of Technology, Nagano 381-8550, Japan.}
\altaffiltext{25}{Jodrell Bank Centre for Astrophysics, The University of Manchester, 
Manchester, M13 9PL, UK.}
\altaffiltext{26}{Tokyo Metropolitan College of Industrial Technology, Tokyo 116-8523, Japan.}
\altaffiltext{27}{Advanced Space Technology Research Group, Institute of
Aerospace Technology, Japan Aerospace Exploration Agency (JAXA), Tokyo,
Japan}
\altaffiltext{28}{Department of Physics, University of Auckland, Private Bag 92-019, 
Auckland 1001, New Zealand.}
\altaffiltext{29}{Institut d'Astrophysique de Paris, 98bis Boulevard Arago, 75014
Paris, France.}
\altaffiltext{30}{European Southern Observatory, Casilla 19001, Vitacura 19, 
Santiago, Chile.}
\altaffiltext{31}{Department. of Physics / Boyden Observatory, University of the Free
State, Bloemfontein 9300, South Africa.}
\altaffiltext{32}{Astronomisches Rechen-Institut, Zentrum f\"ur~Astronomie,
Heidelberg University, M\"{o}nchhofstr.~12--14, 69120 Heidelberg, Germany}
\altaffiltext{33}{Lawrence Livermore National Laboratory, IGPP, P.O. Box 808,
Livermore, CA 94551, USA}
\altaffiltext{34}{University of Tasmania, School of Maths and Physics, Private
bag 37, GPO Hobart, Tasmania 7001, Australia}
\altaffiltext{35}{Department of Physics, University of Rijeka, Omladinska 14, 51000 Rijeka, Croatia}
\altaffiltext{36}{Technical University of Vienna, Dept. of Computing, Wiedner
Hauptstrasse 10, Vienna, Austria}
\altaffiltext{37}{LATT, Universit\'e de Toulouse, CNRS, 14 av. E. Belin, 31400 
Toulouse, France}
\altaffiltext{38}{SUPA, University of St Andrews, School of Physics \& Astronomy,
North Haugh, St Andrews, KY16~9SS, United Kingdom}
\altaffiltext{39}{Niels Bohr Institute, Astronomical Observatory, Juliane Maries
Vej 30, DK-2100 Copenhagen, Denmark}
\altaffiltext{40}{Michelson Science Center, California Institute of Technology, MS 100-22, 770 South Wilson Avenue, Pasadena, CA 91125, USA}
\altaffiltext{41}{Perth Observatory, Walnut Road, Bickley, Perth 6076, Australia}
\altaffiltext{42}{South African Astronomical Observatory, P.O. Box 9 Observatory
7935, South Africa}
\altaffiltext{43}{Astrophysics Research Institute, Liverpool John Moores University,
Twelve Quays House, Egerton Wharf, Birkenhead CH41 1LD, UK}
\altaffiltext{44}{Isaac Newton Group of Telescopes, Apartado de Correos 321, 
E-38700 Santa Cruz de la Palma, Canary Islands, Spain}
\altaffiltext{45}{ESO, Karl-Schwarzschild-Strasse 2, D-85748 Garching bei München}
\altaffiltext{46}{LAOG, Observatoire de Grenoble BP 53 F-38041 GRENOBLE, France}

\vfill
\vfill
\vfill

\begin{abstract}
We combine all available information to constrain the nature of
OGLE-2005-BLG-071Lb, the second planet discovered by microlensing
and the first in a high-magnification event.  These include 
photometric and astrometric measurements from {\it Hubble Space Telescope}, 
as well as constraints from higher order effects extracted from the 
ground-based light curve, such as microlens parallax, planetary orbital motion
and finite-source effects.  Our primary analysis leads to the conclusion that 
the host of Jovian planet OGLE-2005-BLG-071Lb is an M dwarf in the foreground disk with mass $M=0.46\pm 0.04\,M_\odot$, distance $D_{l} = 3.2\pm 0.4\,$kpc,
and thick-disk kinematics $v_{\rm LSR}\sim 103\,\rm km\,s^{-1}$.
From the best-fit model, the planet has mass $M_p = 3.8\pm 0.4\,M_{\rm Jupiter}$,
lies at a projected separation $r_\perp = 3.6\pm 0.2\,$AU from its host and so has an equilibrium temperature of $T \sim 55$ K, i.e., similar to Neptune. 
A degenerate model less favored by $\Delta{\chi^2} = 2.1$ (or 2.2, depending on
the sign of the impact parameter)
gives similar planetary mass $M_p = 3.4\pm 0.4\,M_{\rm Jupiter}$ 
with a smaller projected separation, $r_\perp = 2.1\pm 0.1\,$AU, and 
higher equilibrium temperature $T \sim 71$ K.
These results from the primary analysis suggest that OGLE-2005-BLG-071Lb is
likely to be the most massive planet yet discovered that is hosted by an M dwarf.
However, the formation of such high-mass planetary companions in the outer 
regions of M-dwarf planetary systems is predicted to be unlikely within
the core-accretion scenario. 
There are a number of caveats to this primary analysis, which assumes (based on real but limited evidence) 
that the
unlensed light coincident with the source is actually due to the lens, that is,
the planetary host.  However, these caveats could mostly
be resolved by a single astrometric measurement a few years after
the event.
\end{abstract}
 
\keywords{gravitational lensing -- planetary systems -- Galaxy: bulge}
 
\section{Introduction
\label{intro}}
 
Microlensing provides a powerful method to detect extrasolar planets.
Although only six microlens planets have been found to date
\citep{ob03235,ob05071,ob05390,ob05169,ob06109,mb07400}, these include two major
discoveries. First, two of the planets are ``cold Neptunes'',
a high discovery rate in this previously inaccessible region of parameter
space, suggesting this new class of 
extrasolar planets is common \citep{ob05169, ob05390b}.
Second, the discovery of the first Jupiter/Saturn analog via a
very high-magnification event with substantial sensitivity to multiple planets
indicates that solar system analogs may be prevalent among planetary
systems \citep{ob06109}.
Recent improvements in search techniques and future major upgrades should 
increase the discovery rate of microlensing planets substantially \citep{future}.

Routine analysis of planetary microlensing light curves yields the
planet/star mass ratio $q$ and the planet-star projected separation
$d$ (in units of the angular Einstein radius).  However, because the
lens-star mass $M$ cannot be simply extracted from the light curve, 
the planet mass $M_p=q M$ remains, in general, equally uncertain.

The problem of constraining the lens mass $M$ is an old one.  When 
microlensing experiments were initiated in the early 1990s, it was
generally assumed that individual mass measurements would be impossible
and that only statistical estimates of the lens mass scale could
be recovered.  However, \citet{gould92} pointed out that
the mass and lens-source
relative parallax, $\pi_\rel\equiv \pi_l-\pi_s$,
are simply related to two observable
parameters, the angular Einstein radius,
$\theta_\e$, and the Einstein radius projected onto the plane
of the observer, $\tilde r_\e$, 
\begin{equation}
M = {\theta_\e\over\kappa\pi_\e},
\qquad \pi_\rel = \theta_\e\pi_\e.
\label{eqn:observables}
\end{equation}
Here, $\pi_\e= \au/\tilde r_\e$ is the ``microlens parallax'' and
$\kappa\equiv 4G/(c^2\,\au)\sim 8.1\,{\rm mas}/M_\odot$. 
See \citet{gould00b} for an illustrated derivation of these relations.

In principle, $\theta_\e$ can be measured by comparing some structure
in the light curve to a ``standard angular ruler'' on the sky.
The best example is light-curve distortions due to the finite angular
radius of the source $\theta_*$ \citep{gould94}, which usually can be
estimated very well from its color and apparent magnitude \citep{yoo}.
While such finite-source effects are rare for microlensing events
considered as a whole, they are quite common for planetary events.
The reason is simply that the planetary distortions of the light curve
are typically of similar or smaller scale than $\theta_*$.  In fact,
all six planetary events discovered to date show such effects.  
Combining $\theta_\e$ with the (routinely measurable) 
Einstein radius crossing time $t_\e$ yields
the relative proper motion $\mu$ in the geocentric frame,
\begin{equation}
\mu_{\rm geo} = {\theta_\e\over t_\e},
\label{eqn:observables2}
\end{equation}
From equation~(\ref{eqn:observables}), measurement of $\theta_\e$ by itself
fixes the product $M\pi_\rel = \theta_\e^2/\kappa$.  Using priors
on the distribution of lens-source relative parallaxes, 
one can then make a statistical estimate of the lens mass $M$ and
so the planet mass $M_p$. 

To do better, one must develop an additional constraint.  This
could be measurement of the microlens parallax $\pi_\e$, but this
is typically possible only for long events.  Another possibility
is direct detection of the lens, either under the ``glare'' 
of the source during and immediately after the event, or displaced from 
the source well after the event is over. 
\citet{bennett06} used the latter technique to
constrain the mass of the first microlensing planet, 
OGLE-2003-BLG-235/MOA-2003-BLG-53Lb.  They obtained 
{\it Hubble Space Telescope ({\it HST})} Advanced Camera for Surveys (ACS)
images in $B$, $V$, and $I$ at an epoch $\Delta t=1.78$ years after the event. 
They found astrometric offsets of the (still overlapping) lens and source light
among these images of up to $0.7\,\rm mas$.  Knowing the lens-source
angular separation $\Delta\theta=\mu\Delta t$ from the already
determined values of $\theta_\e$ and $t_\e$, they were able to
use these centroid offsets to fix the color and magnitude of the
lens and so (assuming that it was a main-sequence star) its mass.

While the planet mass is usually considered to be the most 
important parameter that is not routinely derivable from the light
curve, the same degeneracy impacts two other quantities as well,
the distance and the transverse velocity of the lens.  Knowledge
of these quantities could help constrain the nature of the lens,
that is, whether it belongs to bulge, the foreground disk, or possibly
the thick disk or even the stellar halo.  Since microlensing
is the only method currently capable of detecting planets in populations well
beyond the solar neighborhood, extracting such information would
be quite useful.  Because the mass, distance, and transverse
velocity are all affected by a common degeneracy, constraints
on one quantity are simultaneously constraints on the  others.
As mentioned above, simultaneous measurements of $\theta_\e$ and $\pi_\e$
directly yield the mass. However, clearly from equation~(\ref{eqn:observables})
they also yield the distance, and hence (from eq.~[\ref{eqn:observables2}])
also the transverse velocity. 
Here, we assemble all available data to constrain the mass, distance
and transverse velocity of
the second microlensing planet, OGLE-2005-BLG-071Lb, whose
discovery we previously reported (\citealt{ob05071}, hereafter Paper I).

\section{Overview of Data and Types of Constraints
\label{sec:overview}}

\begin{figure}
\plotone{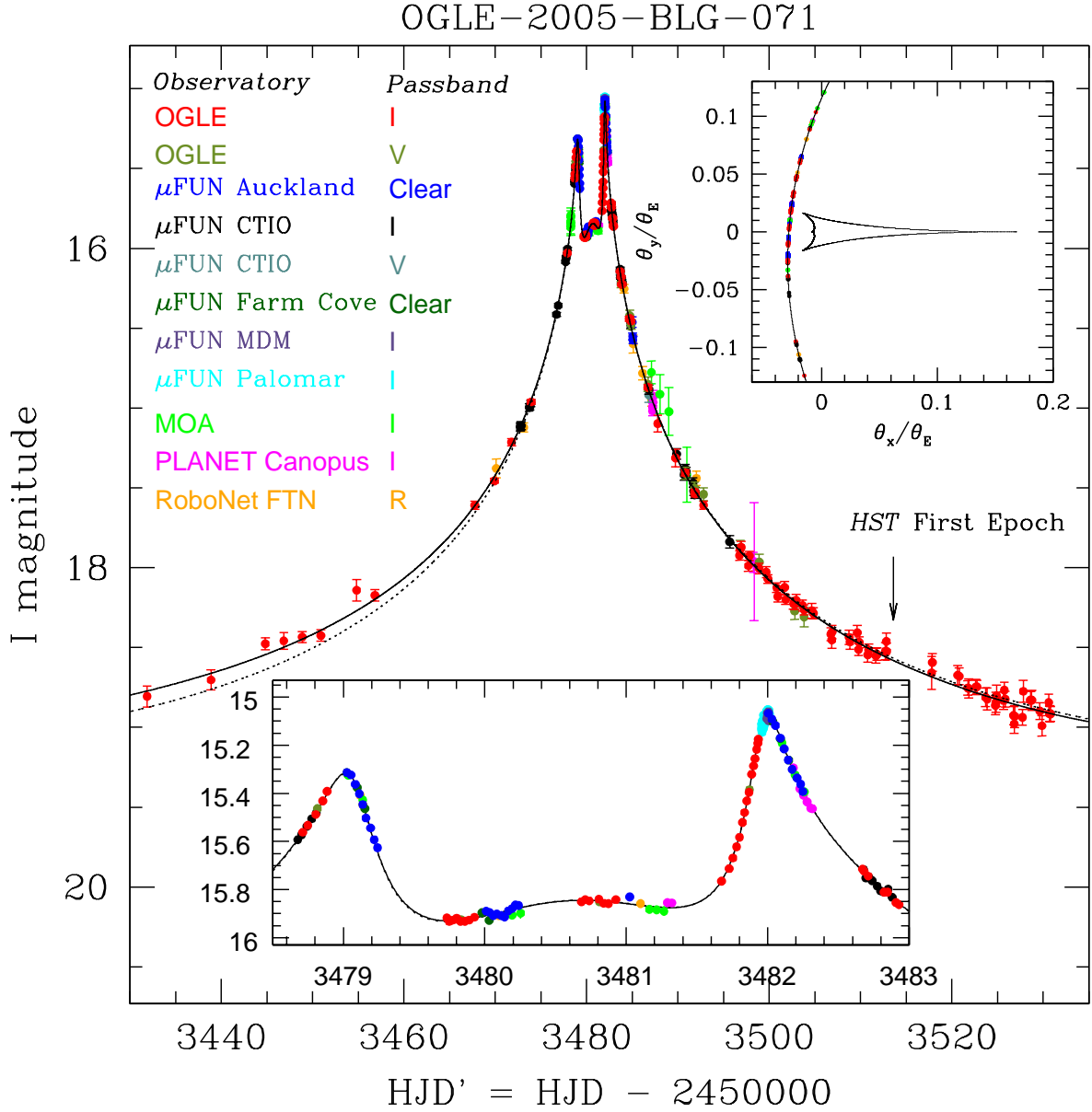}
\caption{Main panel: all available ground-based
data of the microlensing event OGLE-2005-BLG-071. {\it HST} ACS HRC
observations in F814W and F555W were taken at two epochs, once when
the source was magnified by $A\sim2$ ({\it arrow}), and again at 
${\rm HJD} = 2453788.2$ (at baseline). Planetary models that
include ({\it solid}) and excludes ({\it dotted}) microlens parallax
are shown. Zoom at bottom: triple-peak feature that reveals the presence 
of the planet. Each
of the three peaks corresponds to the source passing by a cusp 
of the central caustic induced by the planet. Upper inset: 
trajectory of the source relative to the lens system in the units
of angular Einstein radius $\theta_{\e}$. The lens star is at $(0,0)$, 
and the star-planet axis is parallel to the x-axis. The best-fit angular
size of the source star in units of $\theta_{\e}$ is $\rho \sim 0.0006$, 
too small to be resolved in this figure.}
\label{fig:lc}
\end{figure}

The light curve consists of 1398 data points from 
9 ground-based observatories (see Fig.~\ref{fig:lc}), 
plus two epochs of {\it HST} ACS data in the F814W ($I$) and 
F555W ($V$) filters.
The primary ground-based addition relative to Paper I is 
late-time data from OGLE, which continued to monitor the event down to baseline
until HJD $=2453790.9$.

These data potentially provide constraints on several parameters in
addition to those reported in Paper I. First, the light curve
shows a clear asymmetry between the rising and falling parts of the light
curve, which is a natural result of microlens parallax due to the Earth's 
accelerated motion around the Sun (see the best-fit model 
without parallax effects plotted in dotted line in Fig.~\ref{fig:lc}).
However, it is important to keep in mind that such distortions are equally well
produced by ``xallarap'' due to accelerated motion of the source
around a companion. \citet{poindexter05} showed that it can be difficult 
to distinguish between the two when, as in the present case, the effect 
is detected at $\Delta{\chi^2} \simle 100$.

Second, the two pronounced peaks of the light curve, 
which are due to ``cusp approaches'' (see the bottom inset of Fig.~\ref{fig:lc}), 
are relatively sharp and have good coverage. These peaks would tend to be 
``rounded out'' by finite-source effects, so in principle it may be possible to
measure $\rho$ (i.e., $\theta_*$ in the units of $\theta_\e$) 
from these distortions.

Third, the orbital motion of the planet can give rise to two effects: 
rotation of the caustic about the center of the mass
and distortion of the caustic due to expansion/contraction of the 
planet-star axis. The first changes
the orientation of the caustic structure as the event evolves while
the second changes its shape. These effects are expected to be quite
subtle because the orbital period is expected to be of an order of 10 years
while the source probes the caustic structure for only about 4 days.
Nevertheless, they can be very important for the interpretation of the
event.

Finally, the {\it HST} data cover two epochs, one at 23 May 2005 
(indicated by the arrow in Fig.~\ref{fig:lc})
when the magnification was about $A=2$  and the other 
at 21 Feb 2006 when the event was very nearly at baseline, $A\sim1$.  These data
could potentially yield four types of information.  First, they
can effectively determine whether the blended light is ``associated''
with the event or not. The blended light is composed of sources in the same photometric
aperture as the magnified source, but that do not become magnified
during the event.
If this light is due to the lens, a companion
to the lens, or a companion to the source, it
should fall well within the ACS point spread function (PSF) of the source.
On the other hand, if it is due to a
random interloper along the line of sight, then it should be
separately resolved by the ACS or at least give rise to a distorted PSF.
Second, the {\it HST} data can greatly improve the estimate of the color of 
the blended light.  The original model determined the source
fluxes in both OGLE $V$ and $I$ very well, and of course the baseline
fluxes are also quite well determined. So it would seem that the blended fluxes, which
are the differences between these two, would also be well determined.
This proves to be the case in the $I$ band.  However, while the source
flux is derivable solely from flux differences over the light curve
(and so is well determined from OGLE difference image analysis -- DIA
-- \citealt{wozniak}), the baseline flux depends critically on the zero point
of PSF-fitting photometry, whose accuracy is fundamentally limited in
very crowded bulge fields.  The small zero-point errors turn out 
to have no practical impact for the relatively bright $I$ background light,
but are important for the $V$ band.  Third, one might hope to measure
a centroid shift between the two colors in the manner of \citet{bennett06}. 
Last, one can derive the source proper motion $\bdv{\mu}_s$
from {\it HST} data (at least relative to the mean motion of bulge
stars).  This is important, because the event itself yields the
source-lens {\it relative} proper motion, $\bdv{\mu}_{\rm geo}$.
Hence, precise determination of $\bdv{\mu}_{l}$ requires
knowledge of two proper-motion differences, first 
the heliocentric proper motion 
\begin{equation}
\bdv{\mu}_{\rm hel} = \bdv{\mu}_{l} - \bdv{\mu}_{s}, 
\label{eqn:mu_hel}
\end{equation}
and second, the offset between the heliocentric and geocentric proper
motions
\begin{equation}
\bdv{\mu}_{\rm hel} - \bdv{\mu}_{\rm geo} = 
{{\bdv{v}_{\earth} \pi_{\rm rel}} \over {\rm AU}}.
\label{eqn:mu_convert}
\end{equation}
Here, $\bdv{v}_{\earth}$ is the velocity of the Earth relative to the 
Sun at the time of peak magnification $t_0$. Note that, if the lens-source relative parallax 
$\pi_{\rel}$ is known, even approximately, then the latter difference can be 
determined quite well, since its {\it total} magnitude is just
$0.6\,\masyr (\pi_\rel/0.17\,\mas)$.

\section{Constraining the Physical Properties of the Lens and its Planetary
Companion
\label{sec:host}}

In principle, all the effects summarized in \S~\ref{sec:overview} 
could interact with each other and with the parameters previously determined, 
leading potentially to a very complex analysis.  In fact, we will show 
that most effects can be treated as isolated from one another, which greatly facilitates the exposition. 
In the following sections, we will discuss the 
higher order microlens effects in the order of their impact on the 
ground-based light curve, starting with the strongest, that is, parallax effects 
(\S~\ref{sec:parallax}), followed by planetary orbital motion 
(\S~\ref{sec:orbit}) and finally the weakest, finite-source effects 
(\S~\ref{sec:finitesource}). To study these effects, we implement Markov 
chain Monte Carlo (MCMC) with an adaptive step-size Gaussian sampler 
\citep{mcmc} to perform the model fitting and obtain the uncertainties 
of the parameters. The {\it HST} astrometry is consistent with no
$(V-I)$ color-dependent centroid shift in the first epoch, while such a shift
is seen in the second epoch observations (\S~\ref{sec:astrometry}). 
In addition, the PSF of the source shows no sign of
broadening due to the blend, suggesting that the blend is associated with
the event (\S~\ref{sec:hst}). Therefore, the {\it HST} observations provide 
good evidence that the blend is likely due to the lens. In \S~\ref{sec:astrometry},
it is shown, under such an assumption, how the astrometry can be used in 
conjunction with finite-source and microlens parallax measurements to constrain the
angular Einstein radius and proper motion (\S~\ref{sec:astrometry}).
In \S~\ref{sec:hst}, we discuss using {\it HST} photometric constraints 
in the form of $\chi^2$ penalties to the MCMC runs to extract the color and brightness
of the blend. The results of these runs, which include all higher order effects
of the ground-based light curve and the {\it HST} photometric constraints,
are summarized as ``MCMC A'' in Table \ref{tab:models}.
Subsequently in \S~\ref{sec:final}, by making
the assumption that the blended light seen by {\it HST} is due to the lens, we 
combine all constraints discussed above to obtain physical 
parameters of the lens star and its planet. The corresponding best-fit
model parameters are reported as ``MCMC B'' in Table \ref{tab:models}. 
The results for the physical parameters from these runs are given in
Table \ref{tab:physical}. Finally we discuss some caveats in the analysis 
in \S~\ref{sec:nonlum} and \S~\ref{sec:xallarap}.

\subsection{Microlens Parallax Effects
\label{sec:parallax}}

A point-source static binary-lens model has 
6 ``geometric-model'' parameters: three ``single-lens'' parameters 
($t_0$, $u_0$, $t_{\e}$), where we define the time of ``peak'' magnification
(actually lens-source closest approach) $t_0$ and the impact 
parameter $u_0$ with respect to the center of mass of
the planet-star systems; and three  ``binary-lens'' parameters 
($q$, $d$, $\alpha$), where $\alpha$ is the angle between the star-planet 
axis and the trajectory of the source relative to the lens. 
In addition, flux parameters 
are included to account for light coming from the source star ($F_{s}$) 
and the blend ($F_{b}$) for each dataset. In this paper, we extend the fitting by 
including microlens parallax, orbital motion and finite-source effects. 
Paper I reported that, within the context of the point-source static 
binary-lens models, the best-fit wide-binary ($d>1$) solution is 
preferred by $\Delta{{\chi}^2} = 22$ over the close-binary ($d<1$) solution.
Remarkably, when we take account of parallax, finite-source and orbital
effects, this advantage is no longer as significant. We discuss
the wide/close degeneracy with more detail in \S~\ref{sec:wideclose}.

The microlens parallax effects are parametrized by $\pi_{\e,{\rm E}}$ and 
$\pi_{\e,{\rm N}}$, following the geocentric parallax formalism by \citet{jin02} and \citet{gould04}.
To properly model the parallax effects, we characterize
the ``constant acceleration degeneracy'' \citep{smith} by probing models
with $u_0 \rightarrow -u_0$ and $\alpha \rightarrow -\alpha$. 
We find that all other parameters remain essentially unchanged under this form of degeneracy. 
In the following sections,
if not otherwise specified, parameters from models with positive $u_0$ are adopted.

\begin{figure}
\plotone{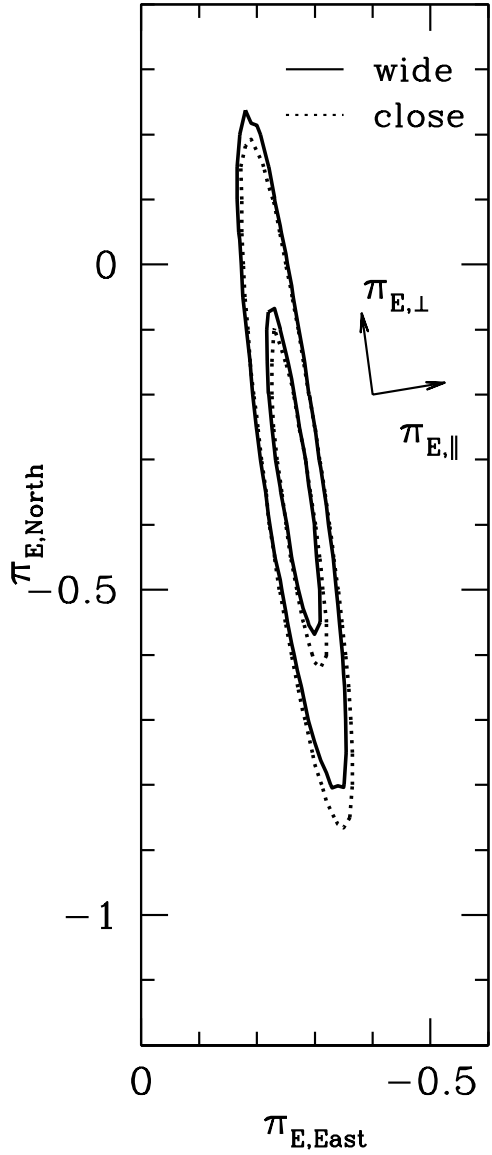}
\caption{
Probability contours ($\Delta{\chi^2}= 1, 4$) of microlens parallax
parameters derived from MCMC simulations
for wide-binary (in solid line) and close-binary (in a dashed line) solutions.
Fig.\ 2 and eq.\ (12) in \citet{gould04} imply that $\pi_{E,\perp}$
is defined so that $\pi_{E,\parallel}$ and $\pi_{E,\perp}$ 
form a right-handed coordinate system.
}
\label{fig:parallax}
\end{figure}

As shown in Figure~\ref{fig:parallax}, microlens parallax is firmly detected in
this event at $> 8 \sigma$ level. Not surprisingly, the error 
ellipse of $\bpi_\e$ is elongated toward $\pi_{\rm E,\perp}$, i.e., 
the direction perpendicular to the position of the Sun at the peak of 
event, projected onto the plane of the sky \citep{thickdisk,poindexter05}.
As a result, $\pi_{\e,{\rm E}}$ is much better determined than $\pi_{\e,{\rm N}}$,

\begin{equation} 
\pi_{\e,{\rm E}} = -0.26 \pm 0.05,\quad \pi_{\e,{\rm N}} = -0.30^{+0.24}_{-0.28} \qquad (\rm wide),
\label{eqn:parallax_wide}
\end{equation}
\begin{equation} 
\pi_{\e,{\rm E}} = -0.27 \pm 0.05,\quad \pi_{\e,{\rm N}} = -0.36^{+0.24}_{-0.27} \qquad (\rm close).
\label{eqn:parallax_close}
\end{equation}

Xallarap (light-curve distortion from reflex motion of the source due to a 
binary companion) could provide an alternate explanation of the detected parallax signals. 
In \S~\ref{sec:xallarap}, we find that the best-fit xallarap parameters are 
consistent with those derived from the Earth's orbit, a result that favors 
the parallax interpretation.

\subsection{Fitting Planetary Orbital Motion
\label{sec:orbit}}
To model orbital motion, we adopt the simplest possible model, with uniform 
expansion rate $\dot{b}$ in binary separation $b$ and uniform binary 
rotation rate $\omega$. Because orbital effects are operative only for 
about 4 days, while the orbital period is of order 10 years, this is 
certainly adequate.
Interestingly, the orbital motion is more strongly detected
for the close solutions (at $\simge 5.5 \sigma$ level) than the wide solutions
(at $\sim 3 \sigma$ level), and as a result, it 
significantly lessens the previous preference of the wide solution that 
was found before orbital motion was taken into account. Further 
discussions on planetary orbital 
motion are given in \S~\ref{sec:planetorbit}.

\subsection{Finite-source effects and Other Constraints
on $\theta_{\e}$
\label{sec:thetae}}

\subsubsection{Color-Magnitude Diagram
\label{sec:cmd}}

\begin{figure}
\plotone{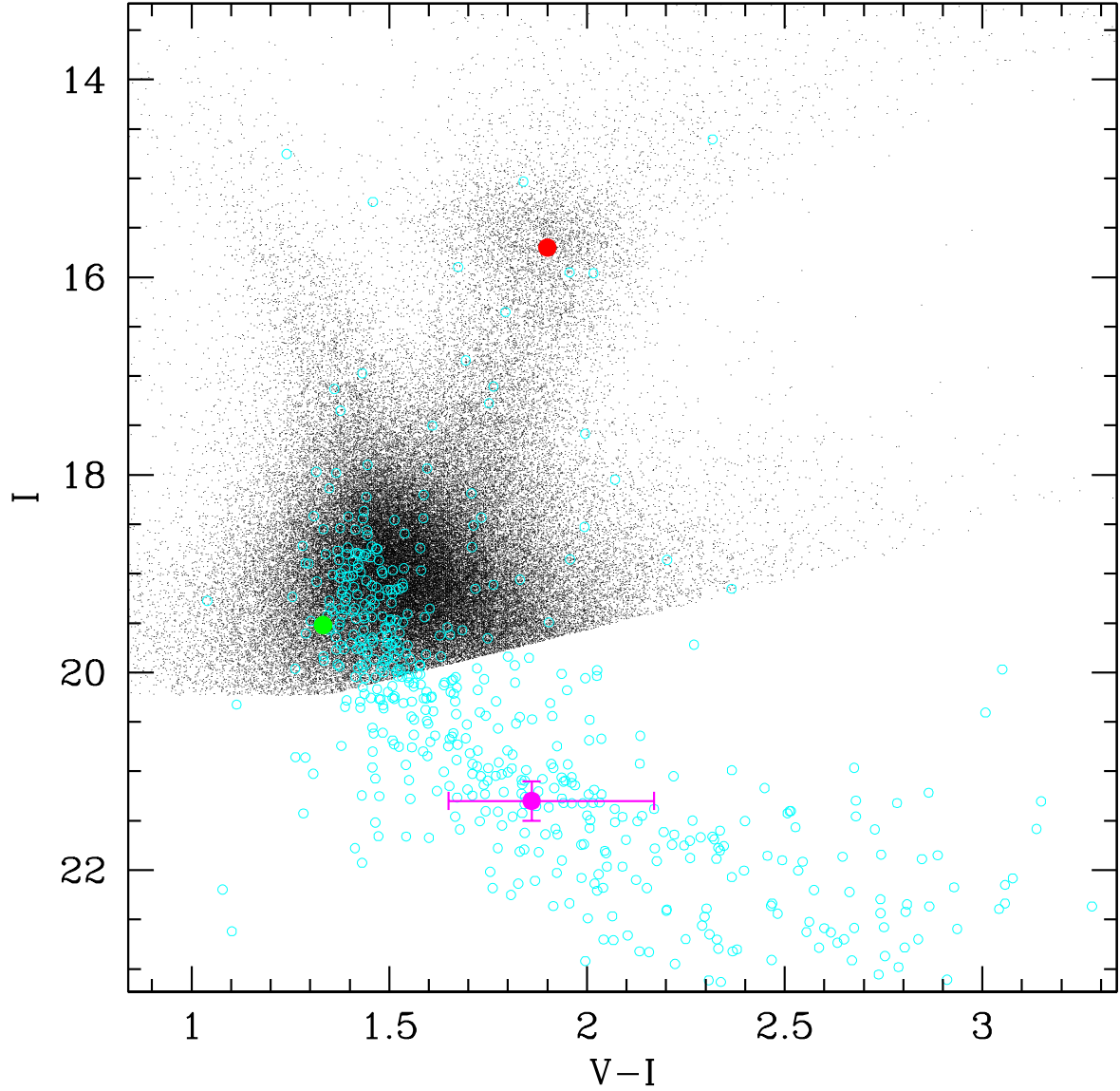}
\caption{CMD for the OGLE-2005-BLG-071 field. Black dots are
the stars with the OGLE $I$-band and $V$-band observations. The red point and
green points show the center of red clump and the source, respectively. The errors
in their fluxes and colors are too small to be visible on the graph. Cyan points
are the stars in the ACS field, which are photometrically aligned with
OGLE stars using 10 common stars. The magenta point with error bars show
the color and magnitude of the blended light.}
\label{fig:cmd}
\end{figure}

We follow the standard procedure to derive dereddened source 
color and magnitude from the color-magnitude diagram (CMD) of the 
observed field. Figure~\ref{fig:cmd} shows the calibrated OGLE CMD 
(black), with the baseline source being displayed as a green point. The 
$V-I$ color of the source can be determined in a model-independent 
way from linear regression of the $I$-band and $V$-band observations. 
The $I$-band magnitude of the source is also precisely determined 
from the microlens model, and it is hardly affected by any higher order 
effects. The center of red clump (red) is at 
$(V-I,I)_{\rm clump} = (1.89, 15.67)$. 
The Galactic coordinates of the source are at 
$(l,b) = (355.58,-3.79)$. Because the Galactic bulge is a bar-like
structure that is inclined relative to the plane of the sky, the red clump density at this sky position peaks
 behind the Galactic center by $0.15$ mag \citep{bar}. 
Hence, we derive $(V-I, I)_{0,\rm clump} = (1.00, 14.47)$, 
by adopting a Galactic distance $R_0 = 8\,\kpc$. We thereby 
obtain the selective and total extinction toward the source 
$[E(V-I), A_I] = (0.89, 1.20)$ and thus $R_{VI} = A_V/E(V-I) = 2.35$. 
The dereddened color and magnitude of the source is
$((V-I), I)_{s,0} = (0.45, 18.31)$. From its dereddened color $(V-I)_0=0.45$, as well as its absolute
magnitude (assuming it is in the bulge) $M_I\sim 3.65$, we conclude
that the source is a main-sequence turnoff star.
Following the 
method of \citet{yoo}, we transform $(V-I)_0 = 0.45$ to 
$(V-K)_0 = 0.93$ \citep{bessel}, and based on the empirical 
relation between the color and surface brightness for subgiant and main-sequence stars \citep{kervella}, 
we obtain the angular size of the source
\begin{equation}
\theta_* = 0.52 \times 10^{0.2 (19.51-I_s)} \pm 0.05 \,\muas,
\label{eqn:thetastar}
\end{equation}
where $I_s$ is the apparent magnitude of the source in the $I$ band. 
Other features on the CMD shown in Figure~\ref{fig:cmd} are further
discussed in \S~\ref{sec:hst}.

\subsubsection{Photometric Systematics of the Auckland Data set
\label{sec:auckland}}

The Auckland data set's excellent coverage over the two peaks 
makes it particularly useful for probing the finite-source 
effects. Unlike the more drastic ``caustics crossings'' that
occur in some events, 
the finite-source effects during ``cusp approaches'' 
are relatively subtle. Hence one must ensure that the
photometry is not affected 
by systematics at the few percent level when
determining $\rho=\theta_*/\theta_\e$. The Auckland photometry 
potentially suffers from two major systematic effects. 

First, the photometry of constant stars reduced by $\mu$FUN's DoPHOT
pipeline are found to show 
sudden ``jumps'' of up to $\sim 10\%$ when the field crossed the 
meridian each night. The signs and amplitudes of the ``jumps'' 
depend on the stars' positions on the CCD. The Auckland 
telescope was on a German equatorial mount, and hence the camera 
underwent a meridian flip. Due to scattered light, the flat-fielded 
images were not uniform in
illumination for point sources, an effect that can be corrected
by making ``superflats'' with photometry of constant stars \citep{superflat}. 
We have constructed such ``superflats'' for each 
night of Auckland observations using 71 bright isolated comparison stars
across the frame. The DoPHOT instrumental magnitude $m_{i,j}$ for 
star $i$ on frame $j$ is modeled by the following equation:
\begin{equation}
m_{i,j} = m_{0,i} - f(x_{i,j},y_{i,j}) - Z_j - fwhm_{i,j}\times s_i
\label{eqn:superflat}
\end{equation}
where $m_{0,i}$ is the corrected magnitude for star $i$, 
$f(x,y)$ is a biquartic illumination correction as a function
of the $(x,y)$ position on the CCD frame with 14 parameters, $Z_j$ is
a zero-point parameter associated with each frame (but with $Z_1$ 
set to be zero), and $s_i$ is a linear correlation coefficient for
the seeing $fwhm_{i,j}$. A least-squares fit that recursively rejects 4-$\sigma$ outliers is performed
to minimize $\chi^2$. The best-fit $f(x,y)$ is dominated
by the linear terms and has small quadratic terms, while its 
cubic and quartic terms are negligible. The resulting 
reduced $\chi^2$ is close to unity, and the ``jumps'' for all 
stars are effectively eliminated. We apply the biquartic
corrections to the images and then reduce the corrected images 
using the DIA pipeline. The resulting DIA photometry of the 
microlens target is essentially identical (at the $\sim 1\%$ level) 
to that from the DIA reductions of the original Auckland images.

As we now show, this is because DIA photometry automatically removes any artifacts
produced by the first- and second-order illumination distortions
if the sources are basically uniformly distributed across the
frame. For the first-order effect, a meridian flip about the
target (which is very close to center of the frame) will induce 
a change in the flux from the source, but it will also
induce a change in the mean flux from all other stars in the 
frame, which for a linear correction will be the same as the 
change in position of the ``center of light'' of the frame
light. If the frame sources are uniformly distributed over the
frame, the ``center of light'' will be the center of the frame,
which is the same position as the source, therefore introducing
no effects. The second-order transformation is even 
under a rotation of 180 degrees, whereas a meridian flip is odd under
this transformation. Hence the flip has no effects at second order.

Second, the Auckland observations were unfiltered. 
The amount of atmospheric extinction differs for stars with
different colors. As shown in Figure~\ref{fig:cmd}, the source is 
much bluer than most of the bright stars in the field, which dominate
the reference image. So the amount of extinction
for the source is different from the average extinction over the 
whole frame. This difference varies as the airmass changes over time 
during the observations.
Coincidentally, the times of the two peaks were both near maximum 
airmass when the ``differential extinction'' effect is expected to be 
the most severe. To 
investigate this effect, we match the isolated stars in the Auckland
frame with CTIO $I$ and $V$ photometry. We identify 33 bright,
reasonably isolated stars with $|(V-I) - (V-I)_s|< 0.25$.  We
obtain a ``light curve'' for each of these stars, using
exactly the same DIA procedure as for the source.  We measure the mean
magnitude of each of the 33 light curves and subtract this value from each
of the 508 points on each light curve, thereby obtaining residuals that are
presumably due primarily to airmass variation.  For each of the 508
epochs, we then take the mean of all of these residuals.  We recursively
remove outliers until all the remaining points are within $3\,\sigma$
of the mean, as defined by the scatter of the remaining points.
Typically, 1 or 2 of the 33 points are removed as outliers.
The deviations are well fitted by a straight line,
\begin{equation}
{d{\rm Mag}\over d Z} =  0.0347 \pm 0.0016
\label{eqn:color}
\end{equation}
where $Z$ is airmass. The sense of the effect is that stars with the
color of the microlensed source are systematically fainter at high
airmass, as expected.  (We also tried fitting the data to a parabola
rather than a line, but the additional [quadratic] parameter was detected
at substantially below $1\,\sigma$.)\ \ Finally, we apply these
 ``differential extinction'' corrections to the ``superflat''-adjusted 
DIA photometry to remove both photometric systematics. 

In general, the finite-source effects depend on the 
limb-darkening profile of the 
source star in the observed passbands. We find below that in this case, 
the impact proves to be extremely weak. Nevertheless, using the matched Auckland and
CTIO stars, we study the difference between Auckland magnitudes and 
$I$-band magnitude as a function of the $V-I$ color. We find the 
 Auckland clear filter is close to the $R$ band.

\subsubsection{Blending in Palomar and MDM Data
\label{sec:MDMblend}}

Palomar data cover only about 80 minutes, but these include
the cresting of the second peak, from which we derive essentially
all information about finite-source effects.  The Palomar data
are sensitive to these effects through their curvature.
The curvature derived from the raw data can be arbitrarily
augmented in the fit (and therefore the finite-source
effects arbitrarily suppressed) by increasing the blending.
In general, the blending at any observatory is constrained
by observations at substantially different magnifications,
typically on different nights.  However, no such constraints
are available for Palomar, since observations were carried
out on only one night.

We therefore set the Palomar blending $f_b = 0.2\,f_s$, that is,
similar to the OGLE blending.  That is, we assume that
the observed flux variation of 9\%, over the Palomar night, actually
reflects a magnification variation of 
$9\%/[1 - f_b/Af_s] = 9\% + 0.026\%$, where $A\sim 70$
is the approximate magnification on that night.  If our estimate
of the blending were in error by of order unity (i.e. either 
$f_b=0$ or $f_b=0.4\,f_s$), then the implied error in the magnification
difference would be 0.026\%, which is more than an order of magnitude
below the measurement errors. Hence, the assumption of fixed blending
does not introduce ``spurious information'' into the fit even at
the $1\,\sigma$ level. MDM data cover the second peak for only 
$\sim 18$ minutes. For consistency, we treat its blending in the 
same way as Palomar, although the practical impact of this data set is an order of magnitude smaller.

\subsubsection{Modeling the Finite-Source Effects
\label{sec:finitesource}}

After careful tests that are described immediately below, we determined
that all finite-source calculations can be carried out to an accuracy
of $10^{-4}$ using the hexadecapole approximation of \citet{hex} 
(see also \citealt{ondrej}).
This sped up calculations by several orders of magnitude.  We began by
conducting MCMC simulations using the ``loop linking'' finite-source code
described in Appendix A of \citet{ob04343}.  From these simulations, we
found the $4.5\,\sigma$ upper bound on the finite-source parameter
$\rho(4.5\,\sigma)=0.001$.  We then examined the differences between
loop-linking (set at ultra-high precision) and hexadecapole for
light curves at this extreme limit and found a maximum difference of
$10^{-4}$.
Based on \citet{claret}, we adopt linear limb-darkening coefficients 
$\Gamma_I = 0.35$ for the I-band observations and $\Gamma_R = 0.43$ 
for the observations performed in the R-band and the clear filters, 
where the local surface brightness is given by $S(\theta) \propto 1-
\Gamma[1-1.5(1-\theta^2/\theta_*^2)^{1/2}]$. 
Ten additional MCMC runs are performed with $\Gamma_I$ and $\Gamma_R$ that 
differ from the above values by $0.1$ or $0.2$. They result
in essentially the same probability distributions of $\rho$. Therefore, the choice of limb-darkening parameters has no effect on the results. The source size is found to be $\rho = 3.9^{+1.8}_{-2.7}$ for the wide solution
and $\rho = 3.1^{+1.7}_{-2.5}$ for the close solution.
Solutions with $\rho > 0.0009$ are ruled out at more than $3 \sigma$.
The angular Einstein radius is given by $\theta_\e = \theta_* / \rho$. Hence, 
the lack of pronounced finite-source effects yields a $3 \sigma$ lower 
limit: $\theta_\e > 0.6\,\mas$. The lens-source relative proper motion 
in the geocentric frame is simply $\mu_{\rm geo} = \theta_\e / t_\e$. 
The posterior probability distributions of $\bdv{\mu}_{\rm geo}$ derived from these MCMC simulations are compared with those derived from astrometry in 
\S~\ref{sec:astrometry}.

\subsection{{\it HST} Astrometry
\label{sec:astrometry}}

\begin{figure}
\plotone{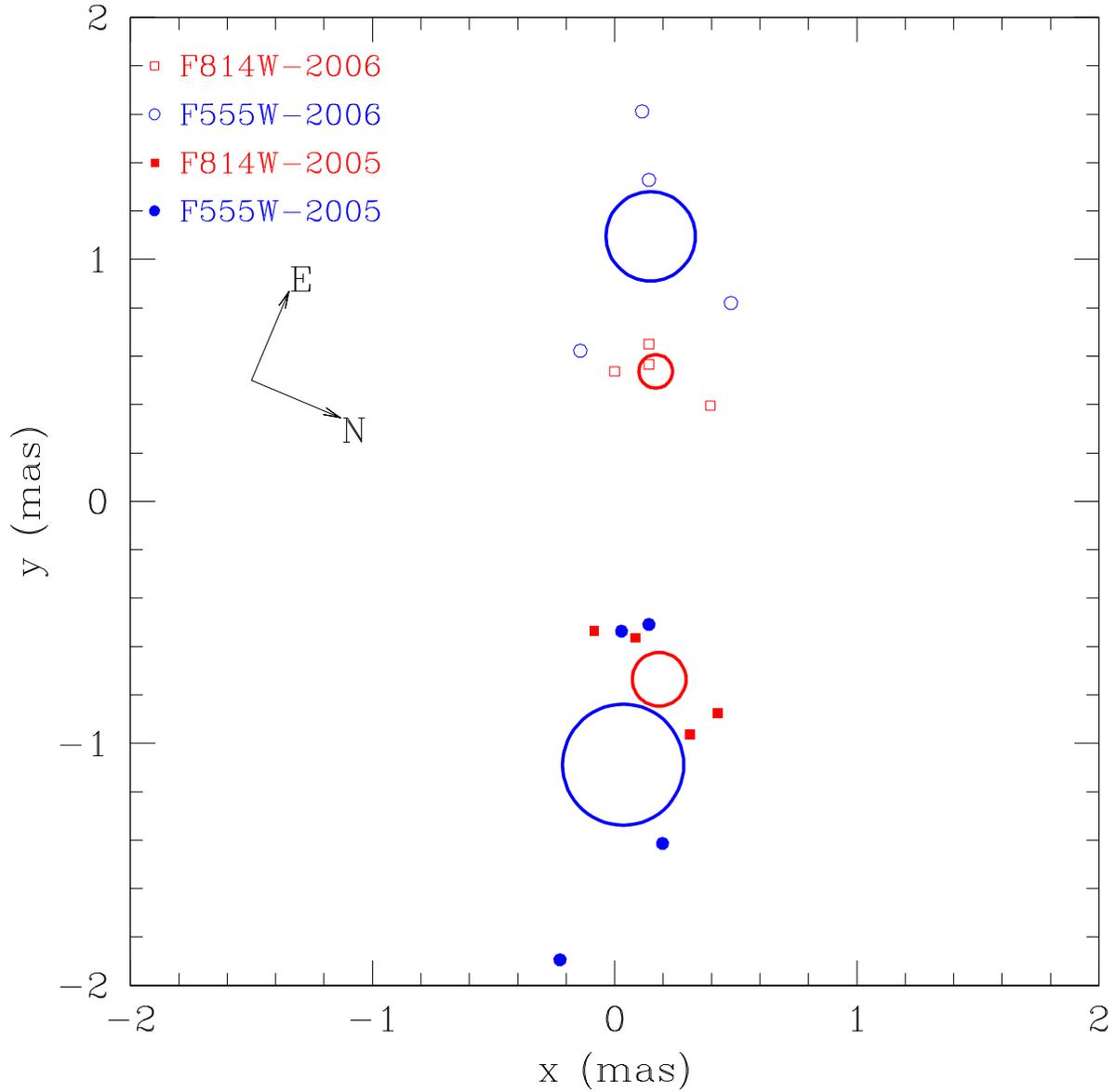}
\caption{{\it HST} ACS astrometric measurements of the target star 
in F814W (red) and F555W (blue) filters in 2005 (filled dots) and 
2006 (open dots). The center positions of the big circles show mean 
values of the 4 dithered observations
in each filter at each epoch while radii of the circle represent
the $1\,\sigma$ errors.
}
\label{fig:hst}
\end{figure}

{\it HST} observations were taken at two epochs (${\rm HJD} = 2453513.6$ 
and ${\rm HJD} = 2453788.2$) with the ACS High Resolution Camera (HRC). 
For each epoch, 4 dithered images were acquired in each of F814W and F555W with
individual exposure times of 225s and 315s respectively. The position of 
the microlens on the {\it HST} frame is in excellent agreement with its 
centroid on the OGLE difference image (within $\sim 0\arcsec.01$). The 
closest star to the source is about $0\arcsec.6$ away. This implies that 
the OGLE photometry of the target star does not contain additional blended light 
that would be identifiable from the
{\it HST} images. Data analysis was carried out using the software
program {\tt img2xym\_HRC} \citep{hrc} in a manner similar to that described in
\citet{bennett06}. Stars are
fitted with an empirical ``library'' PSF that was derived from well-populated
globular cluster fields. These positions
are then corrected with precise distortion-correction models 
(accurate to $\sim 0.01$ pixel). 
We adopted the first F555W frame of the first epoch as the reference frame, 
and used the measured positions of stars in this frame and the frame of
each exposure to define a linear transformation between the exposure
frame and the reference frame. This allowed us to transform the position 
of the target star in each exposure into the reference frame, so that we
see how the target star had moved relative to the other stars. 
The centroid positions of the target star in each
filter and epoch are shown in Figure~\ref{fig:hst}. For convenience, 
in this figure, the positions are displayed relative to the average 
of the centroid positions. The error bars are derived from
the internal scatter of the four dithered images. 
The probability is $P = 38\%$ of measuring the observed separation
(or larger) between F814W and F555W under the assumption that the true offset
is zero. The fact that the blended light is aligned with the source argues
that it is associated with the event (either it is the lens 
itself or a companion to the lens or the source).  We give a more quantitative
statement of this constraint in \S~\ref{sec:hst}.  For the present
we simply note that the $P = 38\%$ probability is 
compatible with the picture that the blend is due to the lens since the first epoch
was only about half of the Einstein-radius crossing time after $t_0$, implying
that the lens-source separation induces only a very small centroid offset,
well below the {\it HST} detection limit.
For the second epoch, the centroid offset is,
\begin{equation}
\Delta{{\bf r}_{{{\rm F814W} - {\rm F555W}}, {\rm East}}}  = -0.52 \pm 0.20 \, {\rm mas},
\qquad \Delta{{\bf r}_{{{\rm F814W} - {\rm F555W}},{\rm North}}} = 0.22 \pm 0.20 \, {\rm mas}.
\label{eqn:second}
\end{equation}
We also calculate the error in the centroid offsets from 
the scatter in such offsets among all comparison stars with
F555W magnitudes within $0.5 \,{\rm mag}$ of the target and find that 
it is consistent with the internally-based error quoted above. 

At the peak of the event, the angular separation between 
the lens and the source was negligible, since $u_0 \ll 1$. We therefore
fix the 
angular positions of the lens and source at a common $\bdv{\theta_{0}}$.
From the CMD (Fig.~\ref{fig:cmd}), most of the stars in the {\it HST} field
are from the bulge. So we set a reference frame that is fixed with 
respect to the bulge field at distance $D_{s}$. The source and lens 
positions at time $t$ are then,
\begin{eqnarray}
\bdv{\theta_{s}}(t) & = & \bdv{\theta_{0}} + \bdv{\mu_{s}}(t-t_0),\nonumber \\
\bdv{\theta_{l}}(t) & = & \bdv{\theta_{0}} + \bdv{\mu_{l}}(t-t_0) + 
\pi_{\rm rel} [{\bdv{s}}(t) - {\bdv{s}}(t_0)],
\label{eqn:astrometry_s_l}
\end{eqnarray}
where ${\bdv{s}}(t)$ is the Earth-to-Sun vector defined by Gould (2004). 
Then by applying equations~(\ref{eqn:mu_hel}) and~(\ref{eqn:mu_convert}), 
the angular separation between the lens and source is,
\begin{eqnarray}
\bdv{\theta_{\rm rel}}(t) = \bdv{\theta_{l}}(t) - \bdv{\theta_{s}}(t) =
\bdv{\mu}_{\rm geo} (t - t_0) + \pi_{\rm rel}\Delta{{\bdv{s}}(t)}
\end{eqnarray} 
where $\Delta{\bdv{s}}(t)$ is given by eq.\ (5) in \citet{gould04}. 

The
centroid of the source images $\bdv \theta^{\prime}_{s}$ is displaced from the source position
by \citep{walker},

\begin{eqnarray}
{\bdv \Delta{\bdv \theta_{s}}(t)} = \bdv{\theta_{s}^{\prime}}(t) - \bdv{\theta_{s}}(t) 
                      = 
{-\bdv{\theta_{\rm rel}}(t)  
\over 
{[{\theta_{\rm rel}(t) / \theta_{\rm E}}]^2 }  + 2
}
\label{eqn:astrometry_b}
\end{eqnarray}

Therefore, one can obtain the centroid position of the lens and the source at time $t$, 
\begin{eqnarray}
\bdv{\theta_{c}}(t) & = & [1-f_l(t)] [\bdv{\theta_{s}}(t) + {\bdv \Delta{\bdv \theta_{s}}(t)}]+ f_l(t) \bdv{\theta_{l}}(t) \nonumber \\
                        & = & \bdv{\theta_{0}} + \bdv{\mu_{s}}(t-t_0) + {\bdv \theta_{\rm rel}(t)[f_l(t) + {1- f_l(t) \over [{\theta_{\rm rel}(t) / \theta_{\rm E}}]^2 + 2}}]
\label{eqn:astrometry_c}
\end{eqnarray}
where $f_l(t)$ is the fraction of the total flux due to the lens.

\begin{figure}
\plotone{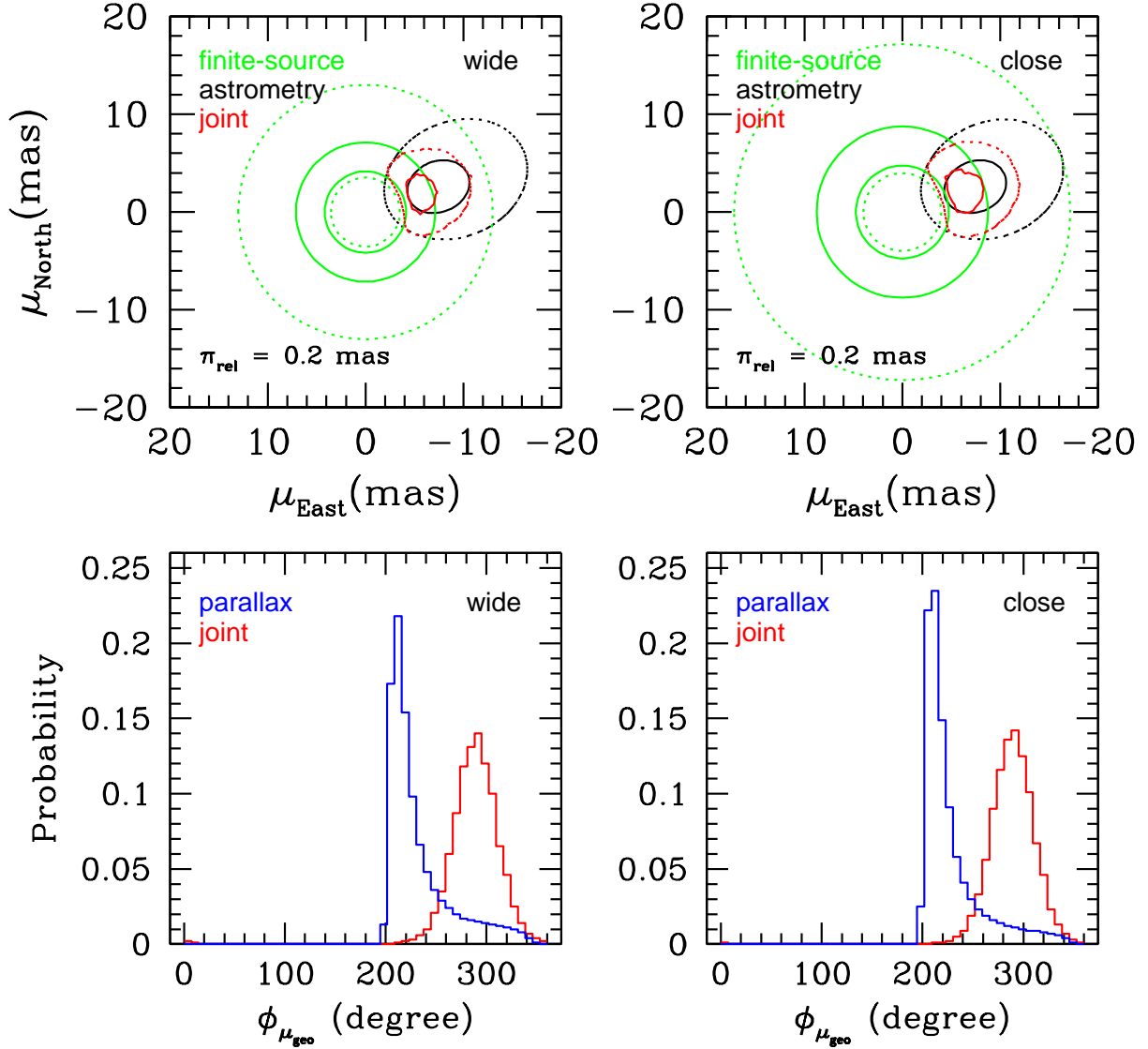}
\caption{
Upper two panels show posterior probability contours at $\Delta{\chi^2}= 1$ 
(solid line) and $4$ (dotted line) for relative lens-source proper motion 
$\bdv{\mu}_{\rm geo}$. 
The left panel is for wide-binary solutions and the right one is for close-binary.
The green contours show the probability distributions constrained by the 
finite-source effects. 
The black contours are derived from {\it HST} astrometry measurements assuming
$\pi_{\rm rel} = 0.2\,\mas$. The red contours show the joint probability
distributions from both constraints. 
The lower two panels show the posterior probability distribution
of the position angle $\phi_{\bdv{\mu}_{\rm geo}}$ of the relative lens-source 
proper motion
for wide-binary and close-binary solutions, respectively. The histogram
in red is derived from the red contours of joint probability for finite source 
and astrometry constraints in the upper panel. The blue histogram represents
that of the microlens parallax. They mildly disagree at $2.5 \sigma$.
}
\label{fig:mu_hst}
\end{figure}

The centroid offset between the two passbands, F814W and F555W, 
is related to the properties of the system by,
\begin{equation}
\Delta{{\bdv{\theta}_c}(t)_{{\rm F814W} - {\rm F555W}}} = [f(t)_{l,{\rm F814W}} - f(t)_{l,{\rm F555W}}]
\times[1 -  {1 \over [{\theta_{\rm rel}(t) / \theta_{\rm E}}]^2 + 2}]{\bdv \theta_{\rm rel}(t)}.
\label{eqn:centroid_color}
\end{equation}
The difference of the blend's fractional flux between F814W and F555W is obtained from ``MCMC A'' described in \S~\ref{sec:hst}.
Consequently, under the assumption that the blend is the lens, we can use 
the measurement of the second-epoch {\it HST} centroid offset to estimate 
the relative proper motion from equation~(\ref{eqn:centroid_color}) for a 
given $\pi_{\rm rel}$.
For purposes of illustration, 
we temporarily adopt $\pi_{\rm rel} = 0.2$ when
calculating the probability distribution of $\bdv{\mu}_{\rm geo}$ (black contours 
in the upper panels of Fig.~\ref{fig:mu_hst}). 
The centroid shift generally favors faster relative proper motion than that 
derived from the source size measurement 
(green contours in Fig.~\ref{fig:mu_hst}), but the difference is 
only at the $\sim 1\sigma$ level. We then get a joint probability 
distribution of $\bdv{\mu}_{\rm geo}$ from both finite-source effects and astrometry, 
which is shown as the red contours in the upper panels of Figure~\ref{fig:mu_hst}.

We then derive the distribution of the $\bdv{\mu}_{\rm geo}$ position angle $\phi_{\bdv{\mu}_{\rm geo}}$ (North 
through East), which is shown by the red histograms in the lower panels 
of Figure~\ref{fig:mu_hst}. Since the direction of the lens-source relative proper 
motion $\bdv{\mu}_{\rm geo}$ is the same as that of the microlens 
parallax $\bdv{\pi}_{\e}$ in the geocentric frame, we have an independent
check on the $\phi_{\bdv{\mu}_{\rm geo}}$ from our parallax measurements, 
whose distribution is plotted as blue histograms in Figure~\ref{fig:mu_hst}. 
Both constraints favor the lens-source proper motion to be generally West,
but they disagree in the North-South component for which both constraints
are weaker. The disagreements
between two histograms is at about $2.5\sigma$ level.

\subsection{``Seeing'' the Blend with {\it HST}
\label{sec:hst}}
If the blend were not the lens (or otherwise associated with the event), 
the PSF of the source would likely be broadened
by the blended star. We examine the {\it HST} F814W images of the target and 45 
nearby stars with similar brightness for each available exposure. We fit 
them with the library PSF produced by \citet{hrc}. 
In order to account for breathing-related changes of focus, 
we fit each of these 45 nearby stars with the library PSF, and construct 
a residual PSF that can be added to the library PSF to produce a PSF that 
is tailor-made for each exposure.
For both epochs, the source-blend combination shows no detectable broadening
relative to the PSFs of other isolated stars in the field.  
From the ground-based light curve, it is already known that $\sim 16\%$ of 
this light comes from the blend. We add simulated stars with the same flux 
as the blended light from 0 to 2.0 pixels away from the
center of the source. We find that the blend would have produced detectable 
broadening of the PSF if it were more than $15\,\mas$ apart from the source 
at the second epoch. Hence, the source-blend separation must then be less 
than about $15\,\mas$.  From the {\it HST} image itself, the density of 
ambient stars at similar magnitudes is $\la 1\,\rm arcsec^{-2}$. 
The probability of a chance interloper is therefore $<0.07\%$, implying that 
the blended light is almost certainly associated with the event, i.e., 
either the lens itself, a companion to the lens, or a companion to the source.
Both of the latter options are further constrained in \S~\ref{sec:nonlum} where,
in particular, we essentially rule out the lens-companion scenario.

As discussed in \S~\ref{sec:overview}, the blended flux in $I$ is
relatively well determined from the ground-based OGLE data alone, but the blended
$V$ flux is poorly determined, primarily because the systematic uncertainty
in the zero point of the baseline flux (determined from PSF fitting)
is of the same order as the blended flux.  Because the {\it HST} image
is very sparse, there is essentially no zero-point error in the
{\it HST} $V$-band flux.  
The problem is how to divide the baseline $V$ flux 
into source and blend fluxes, $F_{\rm base}= F_{s}+F_{b}$.

The standard method of doing this decomposition would be to incorporate the
{\it HST} $V$ light curve into the overall fit, which would
automatically yield the required decomposition.  Since this ``light curve''
consists of two points, the ``fit'' can be expressed analytically
\begin{equation}
F_{s} = {F(t_1) - F(t_2)\over A_1 - 1}, 
\qquad F_{b} = F(t_2) - F_{s}.
\label{eqn:hstfit}
\end{equation}
where we have made the approximation that the second 
observation is at baseline.  Let us then estimate the resulting errors in
$F_{s}$ and $F_{b}$, ignoring for the moment that there is
some uncertainty in $A_1$ due to uncertainties in the general model.
Each of the individual flux measurement is determined from 4 separate
subexposures, and this permits estimates of the errors from the
respective scatters.  These are $\sigma_1=0.01$ and $\sigma_2=0.03$ 
mag.  Hence, the fractional error in $F_{s}$ is
$(2.5/\ln 10)\sigma(F_{s})/F_{s} =
[\sigma_1^2(A_1+r)^2 + \sigma_2^2(1+r)^2]^{1/2}/(A_1 - 1)$, where
$r\equiv F_{b}/F_{s}$.
Adopting, for purposes of illustration, $A_1=2$ and $r=0.1$, this
implies an error $\sigma(V_{{s},HST})$ of 0.04 mag.  This may not
seem very large, but after the subtraction in equation~(\ref{eqn:hstfit}),
it implies an error $\sigma(V_{{b}, HST}) \sim 
\sigma(V_{{s},HST})/r \sim 0.4$ mag. And taking into account of
the uncertainties introduced by model fitting in determining the 
magnifications, the error is expected to be even larger. Hence, we 
undertake an alternate approach.

Because the {\it HST} and OGLE $V$ filters have very nearly the same
wavelength center, $V_{{s}, HST}$ should be nearly identical to
$V_{s, \rm OGLE}$ up to a possible zero-point offset on their
respective magnitude scales.  Because the OGLE data contain many more
points during the event, some at much higher magnification than the
single {\it HST} event point, $V_{s, \rm OGLE}$ is determined
extremely well (for fixed microlensing model), much better than the
0.04 mag error for  $V_{{s},HST}$.  Thus, if the zero-point offset
between the two systems can be determined to better than 0.04 mag,
this method will be superior. Although the $I$-band blend is much better
measured than the $V$-band blend from the ground-based data, 
for consistency we determine the zero-point offset in $I$ by the same
procedure.

\begin{figure}
\plotone{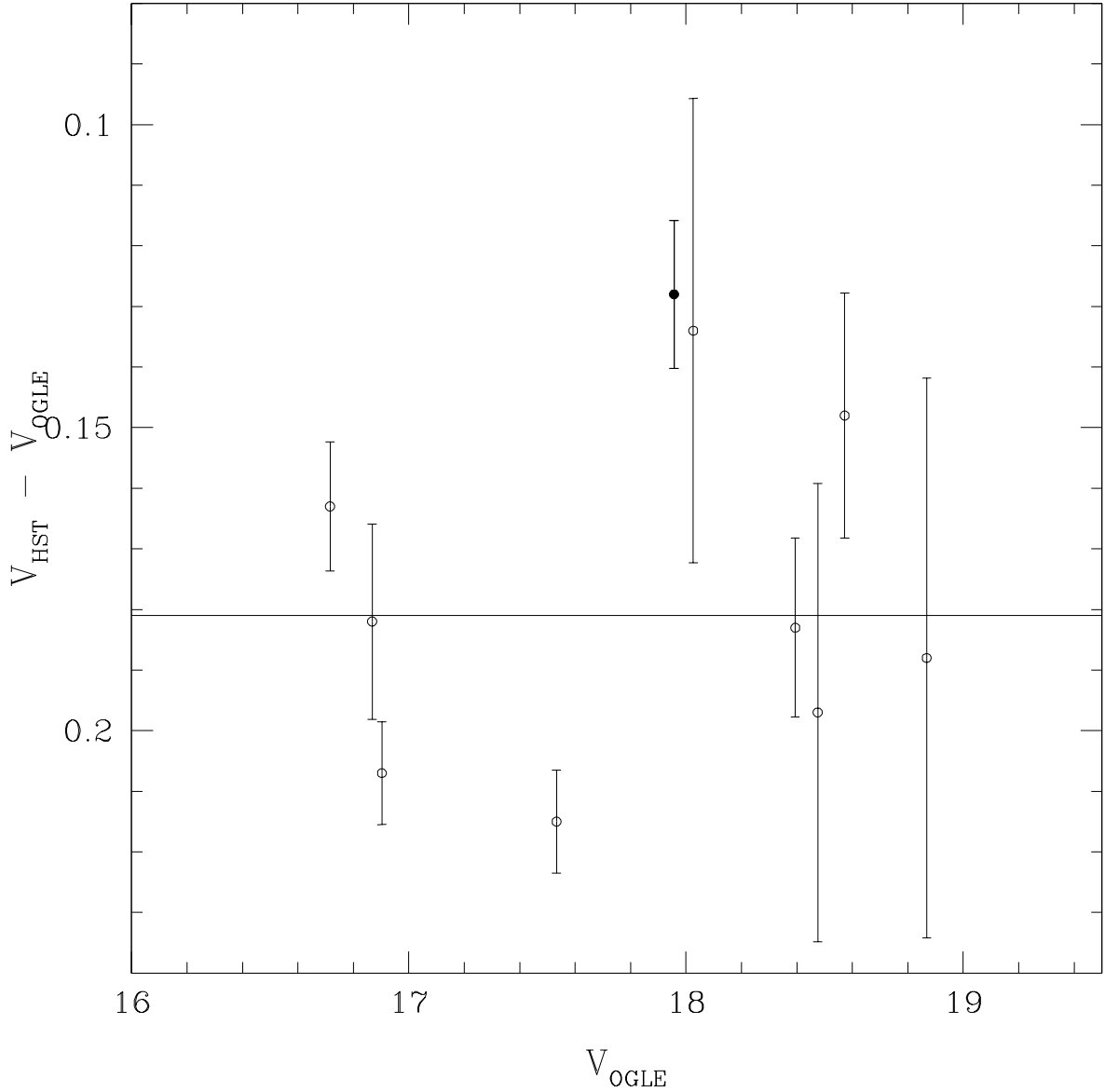}
\caption{Differences between OGLE $V$ and {\it HST} 
F555W magnitudes for the matched stars are plotted against 
their $V$ magnitudes measured by OGLE. To calculate the offset,
we add a $0.017$ mag ``cosmic error'' in quadrature to each point 
in order to reduce $\chi^2/dof$ to unity.
The open circles represent the stars used to establish the final
transformation, and the filled point shows an ``outlier''.
}
\label{fig:hstogle}
\end{figure}

Figure~\ref{fig:hstogle} shows differences between OGLE and {\it HST}
$V$ magnitudes for matched stars in the {\it HST} image.
The error for each star and observatory is determined from the scatter among 
measurements of that star.  We consider only points with $V<19.5$ 
because at fainter magnitudes the scatter grows considerably.  Each star
was inspected on the {\it HST} images, and those that would be
significantly blended on the OGLE image were eliminated. The remaining points 
are fit to an average offset by adding a ``cosmic error''
in quadrature to the errors shown.  We carry out this calculation twice, 
once including the ``outlier'' (shown as a filled circle) and once with 
this object excluded. For the $V$ band, we find offsets of 
$V_{HST} - V_{\rm OGLE}= 0.17 \pm 0.01$ and $0.18 \pm 0.01$, respectively. 
We adopt the following the $V$-band offset
\begin{equation}
\Delta{V} = V_{HST} - V_{\rm OGLE}= 0.18 \pm 0.01.
\label{eqn:hstoglev}
\end{equation}

A similar analysis of the $I$ band leads to
\begin{equation}
\Delta{I} = I_{HST} - I_{\rm OGLE}= 0.08 \pm 0.01.
\label{eqn:hstoglei}
\end{equation}
We find no obvious color terms for either the $V$-band
or $I$-band transformations. As a check, we perform linear regression
to compare the OGLE and {\it HST} $(V-I)$ colors, and we find they
agree within 0.01 mag, which further confirms the color terms are
unlikely to be significant in the above transformations.

We proceed as follows to make {\it HST}-based MCMC (``MCMC A'') estimates 
of $V_{b, {\rm OGLE}}$ and $I_{b, {\rm OGLE}}$ that place the blending 
star on the OGLE-based CMD. 
Since flux parameters are linear, they are often left free and fitted 
by linear least-squares minimization, which significantly accelerates 
the computations. However, for ``MCMC A'', the source fluxes from OGLE 
and {\it HST} 
are treated as independent MCMC parameters so that they can 
help align the two photometric systems as described below. Since {\it HST}
blended light is not affected by light from ambient stars (as OGLE is),
we also leave {\it HST} blended fluxes as independent.
Therefore, in ``MCMC A'', we include the following independent MCMC 
flux parameters,
$F_{I,s, {\rm OGLE}}$, $F_{V,{s}, {\rm OGLE}}$, 
$F_{I,s, {HST}}$, $F_{V,s, {\it HST}}$, 
$F_{I,b, {HST}}$, and $F_{V,{b}, {\it HST}}$, 
which for convenience we express here as magnitudes.  For each model on 
the chain, we add to the light-curve based $\chi^2$ two additional terms
$\Delta\chi^2_{V} = (V_{{s},HST}-V_{s,\rm OGLE}-\Delta V)^2/
[\sigma(\Delta V)]^2$ and 
$\Delta\chi^2_{I} = (I_{{s},{\it HST}}-I_{s,{\rm OGLE}}-\Delta I)^2/ 
[\sigma(\Delta I)]^2$ to enforce the measured offset between the two
systems.
Finally, we evaluate the $V$-band blended flux from {\it HST} and 
convert it to OGLE system, $V_{b, {\rm OGLE}/{\it HST}}= 
V_{s, {\rm OGLE}}-V_{s,{\it HST}}+V_{{b},{\it HST}}$ (and similarly
for $I$ band), where all three terms on the rhs are the individual Monte Carlo realizations of the respective parameters.

The result is shown in Figure~\ref{fig:cmd}, in which the
blend (magenta) is placed on the OGLE CMD.  Also shown, in cyan points, is
{\it HST} photometry (aligned to the OGLE system) of the stars in the
ACS subfield of the OGLE field.  Although this field is much smaller,
its stars trace the main sequence to much fainter magnitudes.
The blend falls well within the bulge main sequence revealed by the
{\it HST} stars on the CMD, so naively the blend can be 
interpreted as being in the bulge. Hence, this diagram is, in itself, most 
simply explained by the blend being a bulge lens or a binary companion of 
the source. However, the measurement of $V-I$ color has
relatively large uncertainty, and it is also consistent with the blend
being the lens (or a companion to it) several kpc in front of the bulge,
provided the blend is somewhat redder than indicated by the best-fit
value of its color. In the following section, we assume the blended light 
seen by {\it HST} is the foreground lens star, and the {\it HST} photometry 
is combined with other information to put constraints on the lens star 
under this assumption.

\subsection{Final Physical Constraints on the Lens and 
Planet}
\label{sec:final}

\subsubsection{Constraints on a Luminous Lens}
\label{sec:finalresults}
In the foregoing, we have discussed two types of constraints on the 
host star properties: the first class of constraints, consisting of 
independent measurements of $\bdv{\pi_\e}$, $\theta_\e$ and $\bdv{\mu}$, 
relate the microlens parameters to the physical parameters of the lens;
the second class are {\it HST} and ground-based observations that 
determine the photometric properties
of the blend. 

\begin{figure}
\plotone{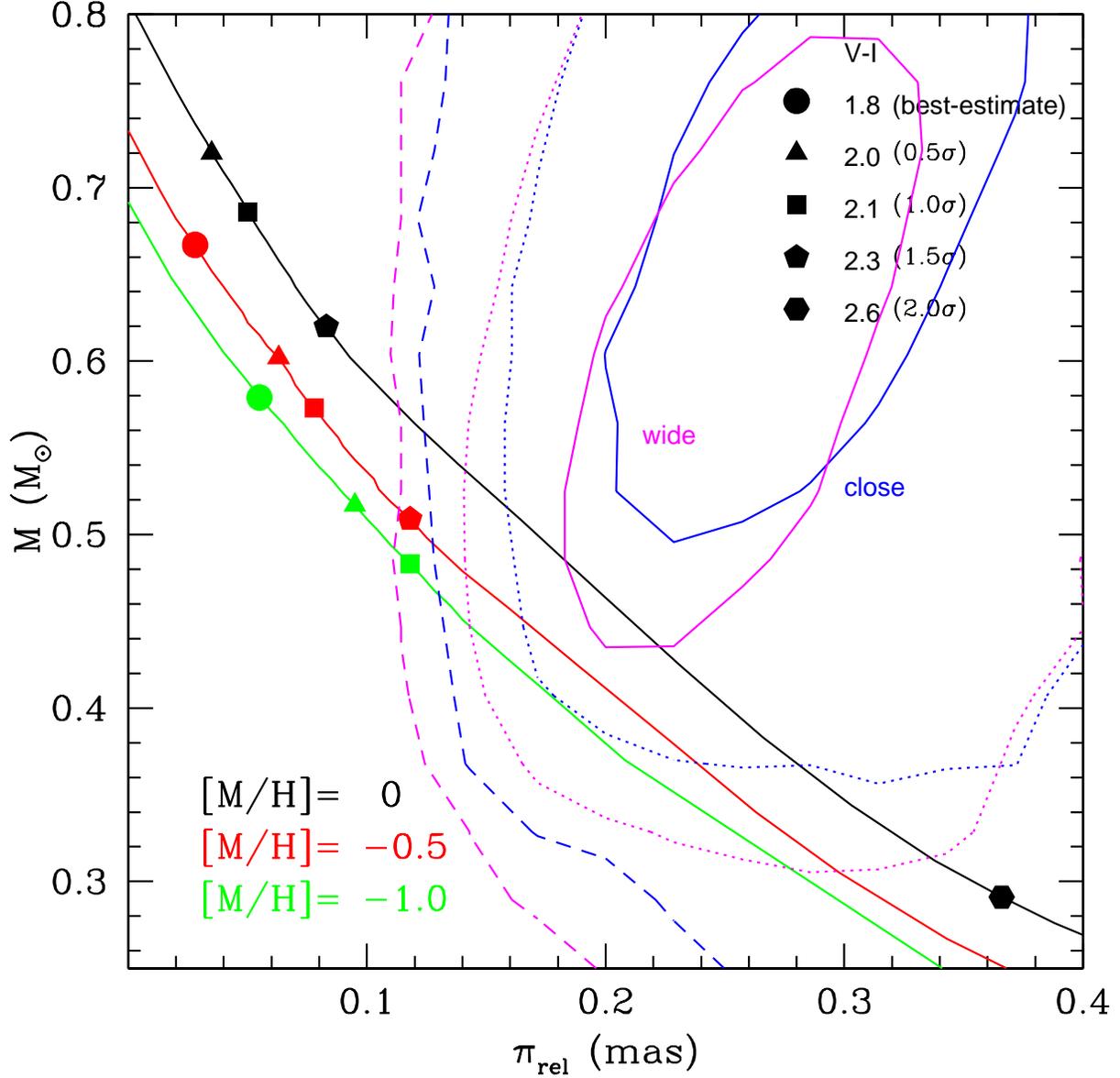}
\caption{Posterior probability distribution of lens 
mass $M$ and relative lens-source parallax $\pi_{\rel}$ from 
MCMC simulations discussed in \S~\ref{sec:finalresults}. The constraints 
include those from parallax effects, finite-source effects and 
relative proper motion measurements from {\it HST} astrometry. 
The $\Delta{\chi^2} = 1, 4, 9$ contours are displayed
in solid, dotted and dashed lines, respectively. Both wide-binary (magenta)
and close-binary (blue) solutions are shown. The lines in black, red
and green represent the predicted $M$ and $\pi_{\rm rel}$ from the isochrones 
for different metal abundances: $\rm [M/H] = 0$ (black), $-0.5$ (red), 
$-1.0$ (green). The points on these lines
correspond to the observed $I$-band magnitude $I = 21.3$ and 
various $V-I$ values $V-I = 1.8\,({\rm best\, estimate}$, filled dots), 
$2.0\,(0.5\,\sigma$, filled triangle), $2.1\,(1.0\,\sigma$, filled
squares), $2.3\,(1.5\,\sigma$, filled pentagons), and $2.6\,(2.0\,\sigma$, 
filled hexagons)}
\label{fig:mass_pirel}
\end{figure}

In this section, we first describe a new set of 
MCMC simulations taking all these constraints into account. 
Similarly to what is done to include {\it HST} photometry in the ``MCMC A''
(see \S~\ref{sec:hst}), we incorporate {\it HST} astrometry constraints
by adding $\chi^2$ penalties to the fittings.
For a given set of microlens parameters, we can derive the physical 
parameters, namely, $M$, $\pi_{\rm rel}$, $\bdv{\mu}_{\rm geo}$, and so 
calculate $\rho = \theta_*/\theta_{\e}$ (from eq.~[\ref{eqn:observables}]) 
and the F814W $-$ F555W centroid offset (from eq.~[\ref{eqn:centroid_color}]). 
Then we assign the $\chi^2$ penalties based on the observed centroid 
offset from \S~\ref{sec:astrometry}. In this way, the MCMC simulations 
simultaneously include all microlens constraints on the lens properties. 
The posterior probability distribution of $M$ and $\pi_{\rm rel}$ are 
plotted in Figure~\ref{fig:mass_pirel}. The $\pi_{\rm rel}$ determination 
very strongly excludes a bulge ($\pi_{\rm rel} \lesssim 0.05$) lens. Note that
by incorporating {\it HST} astrometry, we implicitly assume
that the blend is the lens. 

If the blend is indeed the lens itself, we can also estimate its mass and 
distance from the measured color and magnitude of the blend. 
In doing so, we use theoretical stellar isochrones (M. Pinsonneault 2007, private communication) incorporating the color-temperature 
relation by \citet{iso1,iso2}. We first use an isochrone that has solar 
metal abundance, with stellar masses ranging from $0.25 M_\odot - 1.0 M_\odot$,
and an age of $4 \,{\rm Gyrs}$. The variation in stellar brightness due 
to stellar age is negligible for our purpose. Extinction is modeled as a 
function of $D_l$ by $d{A_I}/d{D_l} = (0.4\,\kpc^{-1})\exp({-w{D_l}})$, 
where $w$ is set to be $0.31\,\kpc^{-1}$ so that the observed value 
$A_I(8.6\,\kpc) = 1.20$ (as derived from CMD discussed in \S~\ref{sec:cmd}) 
is reproduced. Again, the distance to the source is assumed to be $8.6\,\kpc$, 
implying $\pi_s = 0.116\,$mas, and hence that the lens distance is 
$D_l/\kpc = \mas/(\pi_\rel + \pi_s)$. 
In Figure~\ref{fig:mass_pirel},
we show the lens mass $M$ and relative parallax $\pi_{\rm rel}$ 
derived from the isochrone that correspond to 
the observed $I$-band magnitude $I = 21.3$ in black line
and a series of $V-I$ values $V-I = 1.8\,({\rm best\,estimate})$, 
$2.0\,(0.5\,\sigma)$, $2.1\,(1\,\sigma)$, $2.3\,(1.5\,\sigma)$ and 
$2.6\,(2\,\sigma)$ as black points. The observed color is in modest 
disagreement $< 2 \sigma$ with the mass and distance of the lens at 
solar metallicity. We also show analogous trajectories for ${\rm [M/H]} = -0.5$ (red) and ${\rm [M/H]} = -1.0$ (green). The level of agreement 
changes only very weakly with metallicity.

We then include the isochrone information in a new set of MCMC runs 
(``MCMC B''). 
To do so, the {\it HST} blended fluxes in $I$ and $V$ bands can no longer
be treated as independent MCMC parameters. Instead, based on the isochrone with solar 
metallicity, the lens $V-I$ color and $I$ magnitude 
are predicted at the lens mass and distance determined from MCMC parameters. 
Then the {\it HST} $I$-band and $V$-band fluxes are fixed at the 
predicted values in the fitting for each MCMC realization.

\begin{figure}
\plotone{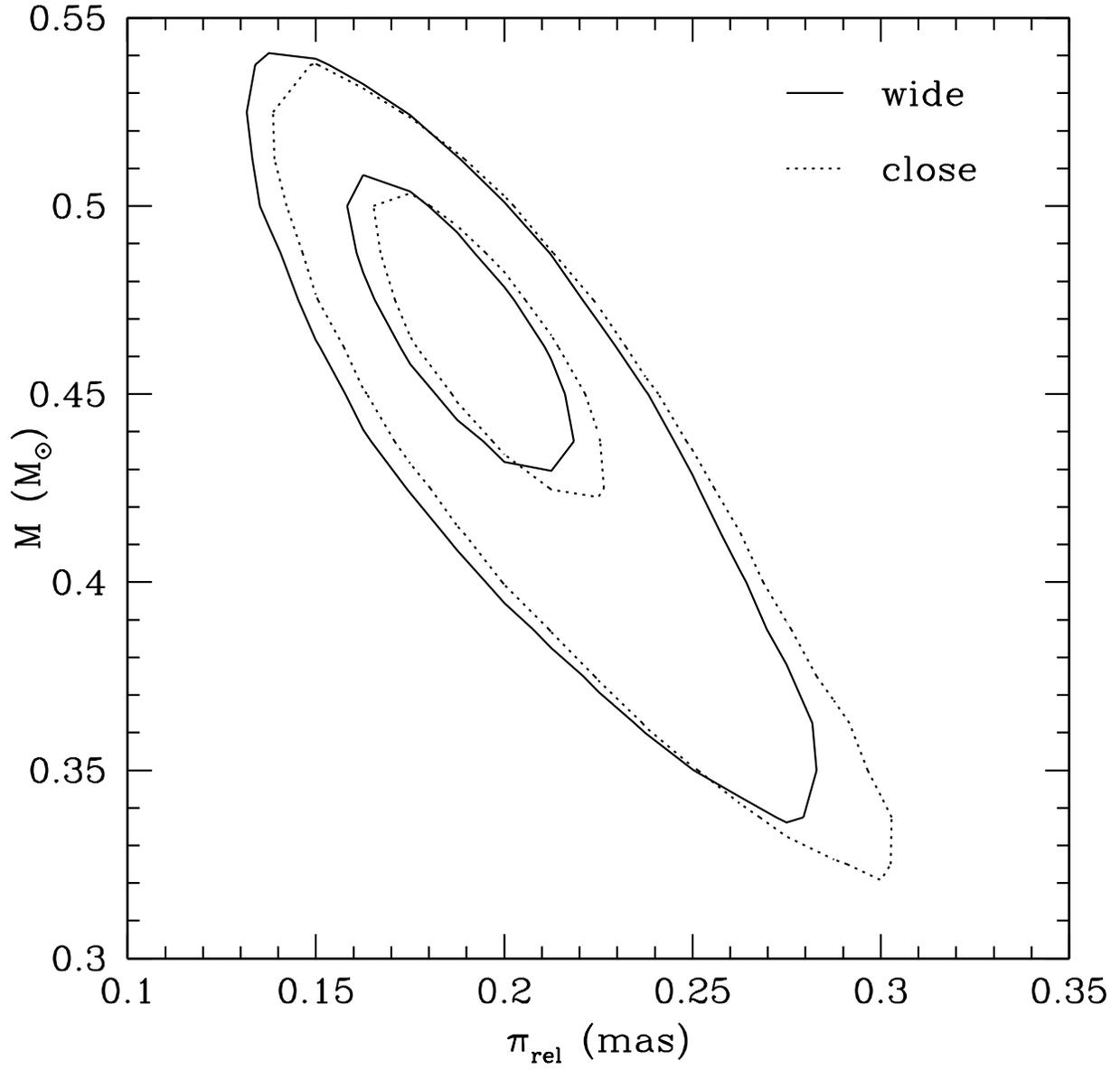}
\caption{
Posterior probability distribution of lens mass $M$ and relative
lens-source parallax $\pi_{\rel}$ from MCMC simulations assuming that the 
blended light comes from the lens star. 
The $\Delta{\chi^2} = 1, 4$ contours are displayed
in a solid line for wide solutions, and in a dotted line for close solutions.
}
\label{fig:mass_pirel2}
\end{figure}

Figure~\ref{fig:mass_pirel2} illustrates 
the constraints on $M$ and $\pi_{\rm rel}$ from the MCMC, which
are essentially the same for both wide-binary (solid contours) 
and close-binary (dashed contours) solutions:
\begin{equation}
M = 0.46 \pm 0.04 \,M_\odot,
\qquad \pi_{\rm rel} = 0.19 \pm 0.03 \,\mas.
\label{eqn:mass_pirel}
\end{equation}
Assuming the source distance at $8.6 \,\kpc$, the $\pi_{\rm rel}$ 
estimates translate to the following lens distance measurement:
\begin{equation}
D_l = 3.2 \pm 0.4 \,\kpc.
\label{eqn:distance}
\end{equation}

Furthermore, we can derive constraints on the planet mass $M_p$ and the 
projected separation between the planet and the lens star $r_\perp$,
\begin{equation}
M_p =  3.8 \pm 0.4 M_{\rm Jupiter},
\qquad r_{\perp} = 3.6 \pm 0.2 {\rm AU} \qquad ({\rm wide}),
\label{eqn:planet_w} 
\end{equation}
and
\begin{equation}
M_p =  3.4 \pm 0.4 M_{\rm Jupiter}, 
\qquad r_{\perp} = 2.1 \pm 0.1 {\rm AU} \qquad ({\rm close}).
\label{eqn:planet_c} 
\end{equation}
The wide solution is slightly preferred over close solution
by $\Delta{\chi^2} = 2.1$.

To examine possible
uncertainties in extinction estimates, we reran our MCMC 
with $A_I$ and $A_V$ that are $10 \%$ higher and lower than the
fiducial values. These runs result in very similar estimates as when
adopting the fiducial values.

From equations~(\ref{eqn:astrometry_c}) and ~(\ref{eqn:astrometry_b}), one can easily obtain the centroid
shift between two epochs in a given passband by ignoring 
${\bdv \Delta{\bdv \theta_{s}}(t)}$ \footnote{
The angular
separations between the source and the lens are $\sim 0.47 \theta_{\e}$ 
and $\sim 4.4 \theta_{\e}$ for the two {\it HST} epochs, respectively. 
Thus the angular position offsets between the centroids of the source 
images and the source are both $\sim 0.21 \theta_{\e}$ and the directions
of the offset relative to the source are almost the same due to the
small impact parameter $u_0$. The difference between lens flux 
fractions of the two epochs are about $7\%$ in $I$ band, so the offsets 
can be confidently ignored in deriving the source proper motion using the 
relative astrometry in F814W at two different epochs.},
\begin{eqnarray}
\bdv{\theta_{c}}(t_2) - \bdv{\theta_{c}}(t_1) & = & \bdv{\mu_{s}}(t_2-t_1) + \bdv{\mu}_{\rm geo} [f_l(t_2)(t_2 - t_0) - f_l(t_1)(t_1 - t_0)] \nonumber\\
&& + \pi_{\rm rel}[f_l(t_2) \Delta{{\bdv{s}}(t_2)} - f_l(t_1) \Delta{{\bdv{s}}(t_1)}]
\label{eqn:centroid_c}
\end{eqnarray}

Because $\bdv{\mu}_{\rm geo}$, $\pi_{\rm rel}$ and $f_l$ in a given
passband can be extracted from the MCMC realizations (``MCMC B''), we can 
use the above equation to measure the source proper motion by making use
of the centroid shift in F814W between two epochs. The source proper motion with 
respect to the mean motion of stars in the {\it HST} field is measured to be
\begin{equation}
\bdv{\mu_{s}} = (\mu_{s,E},\, \mu_{s,N}) = (2.0 \pm 0.2,\,-0.5^{+0.2}_{-0.7})\,{\rm mas}\,{\rm yr}^{-1}.
\label{eqn:mus}
\end{equation}
We obtain similar results with F555W, but with understandably larger errorbars
since the astrometry is more precise for the microlens in F814W.

Combining equations~(\ref{eqn:mu_hel}) and~(\ref{eqn:mu_convert}), the lens proper motion in the heliocentric frame is therefore
\begin{equation}
\bdv{\mu}_{l} = \bdv{\mu}_{\rm geo} + \bdv{\mu}_{s} + 
{{\bdv{v}}_{\earth} \pi_{\rm rel}\over{\rm AU}}.
\label{eqn:helio}
\end{equation}
For each MCMC realization, $\pi_{\rm rel}$ is known, so we can convert the 
lens proper motion to the velocity of the lens in the heliocentric frame 
$\bdv{v}_{l, \rm hel}$ and also in the frame of local standard of rest 
$\bdv{v}_{l, \rm LSR}$ (we ignore the rotation of the galactic bulge). The 
lens velocity in the LSR is estimated to be $v_{l, \rm LSR} = 103 \pm 15\,{\rm km\,s^{-1}}$. This raises
the possibility of the lens being in the thick disk, in which the stars
are typically metal-poor. As shown in Figure~\ref{fig:mass_pirel}, 
the constraints we have cannot resolve the metallicity of the lens star.
\subsubsection{Planetary Orbital Motion
\label{sec:planetorbit}}
\paragraph{Wide/Close Degeneracy
\label{sec:wideclose}}
Binary-lens light curves in general exhibit a well-known ``close-wide'' 
symmetry (\citealt{dominik99, jin05}). Even for some well-covered
caustics-crossing events (e.g., \citealt{macho98smc1}), there are quite
degenerate sets of solutions between wide and close binaries.
In Paper I, we found that the best-fit point-source wide-binary solution
was preferred over close-binary solutions by $\Delta{\chi^2} \sim 22$. But this did
not necessarily mean that the wide-close binary degeneracy was broken, since
the two classes of binaries may be influenced differently by higher order 
effects. 
We find that the $\chi^2$ difference between best-fit wide and 
close solutions is within 1 from ``MCMC A'' and $2.1$ (positive $u_0$) 
or $2.2$ (negative $u_0$) from ``MCMC B''.

However, orbital motion of the planet is subject to additional dynamical
constraints: the projected velocity of the planet should be no greater 
than the escape velocity of the system: 
$v_{\perp} \leq v_{\rm esc}$, where, 
\begin{equation}
v_{\perp} = \sqrt{\dot{d}^2 + (\omega d)^2} {{\rm AU}\over\pi_{l}} \theta_\e,
\label{eqn:proj}
\end{equation}
\begin{equation}
v_{\rm esc} = \sqrt{2 G M \over r} \leq v_{\rm esc,\perp} \equiv 
\sqrt{2 G M \over d \theta_\e {D_{l}}}  = \sqrt{\pi_{l} \over {2 d \pi_\e}}c,
\label{eqn:esc}
\end{equation}
and where $r$ is the instantaneous 3-dimensional planet-star physical separation.
Note that in the last step, we have used equation~(\ref{eqn:observables}).

\begin{figure}
\plotone{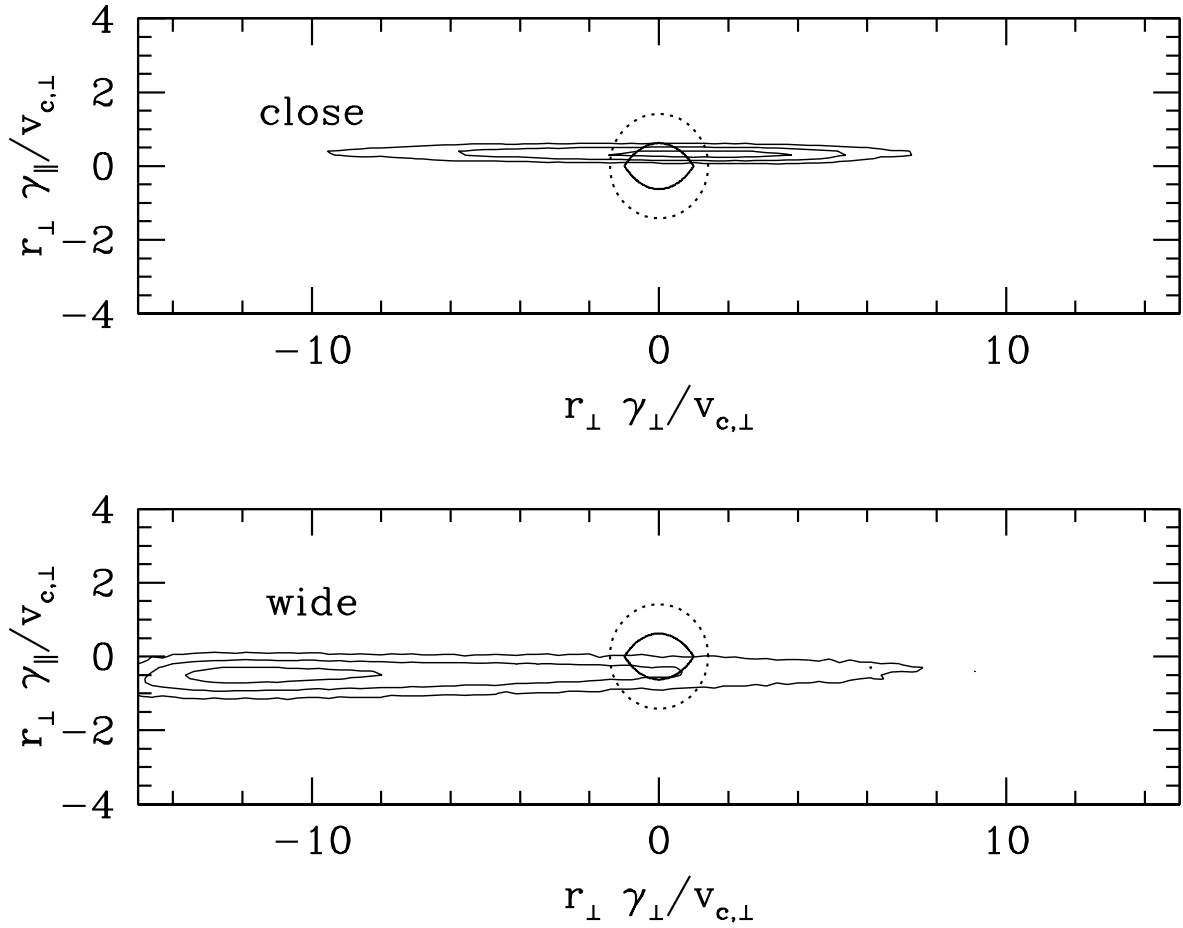}
\caption{Probability contours of projected velocity $r_\perp \bdv{\gamma}$
(defined in Appendix~\ref{sec:appendcirc}) in the units of $v_{\rm c,\perp}$
for both close-binary (upper panel) and wide-binary (lower panel) solutions.
All the solutions that are outside the dotted circle are physically
rejected as the velocities exceed the escape velocity of the system. 
The boundary in a solid line inside the dotted circle encloses the solutions for which circular orbits are allowed.}
\label{fig:rotationplot}
\end{figure}

We then calculate the probability distribution of the ratio
\begin{equation}
\label{eqn:vrat}
{v_\perp^2\over v_{\rm esc,\perp}^2} = 
2{{\rm AU}^2\over c^2}
{d^3[(\dot d/d)^2 + \omega^2] \over [\pi_\e + (\pi_s/\theta_\e)]^3}
{\pi_\e\over\theta_\e}
\end{equation}
for an ensemble of MCMC realizations for both wide and close solutions.
Figure~\ref{fig:rotationplot} shows probability distributions of
the projected velocity $r_{\perp} \bdv{\gamma}$ in the units 
of critical velocity $v_{\rm c,\perp}$, where $r_{\perp} \bdv{\gamma}$ 
is the instantaneous velocity of the planet on the sky, 
which is further discussed in Appendix~\ref{sec:appendcirc} and $
v_{\rm c,\perp} = v_{\rm esc,\perp}/\sqrt{2}$. The dotted circle
encloses the solutions that are allowed by the escape velocity criteria, 
and the solutions that are inside the solid line are consistent
 with circular orbital motion. We find that the best-fit 
 close-binary solutions are physically allowed while the best-fit wide-binary
 solutions are excluded by these physical constraints at $1.6\,\sigma$.
 The physically excluded best-fit wide solutions are favored by 
 $\Delta{\chi^2} = 2.1$ (or 2.2) over the close solutions, so by putting physical
 constraints, the degenerate solutions are statistically not distinguishable 
 at $1\,\sigma$.
 
\begin{figure}
\plotone{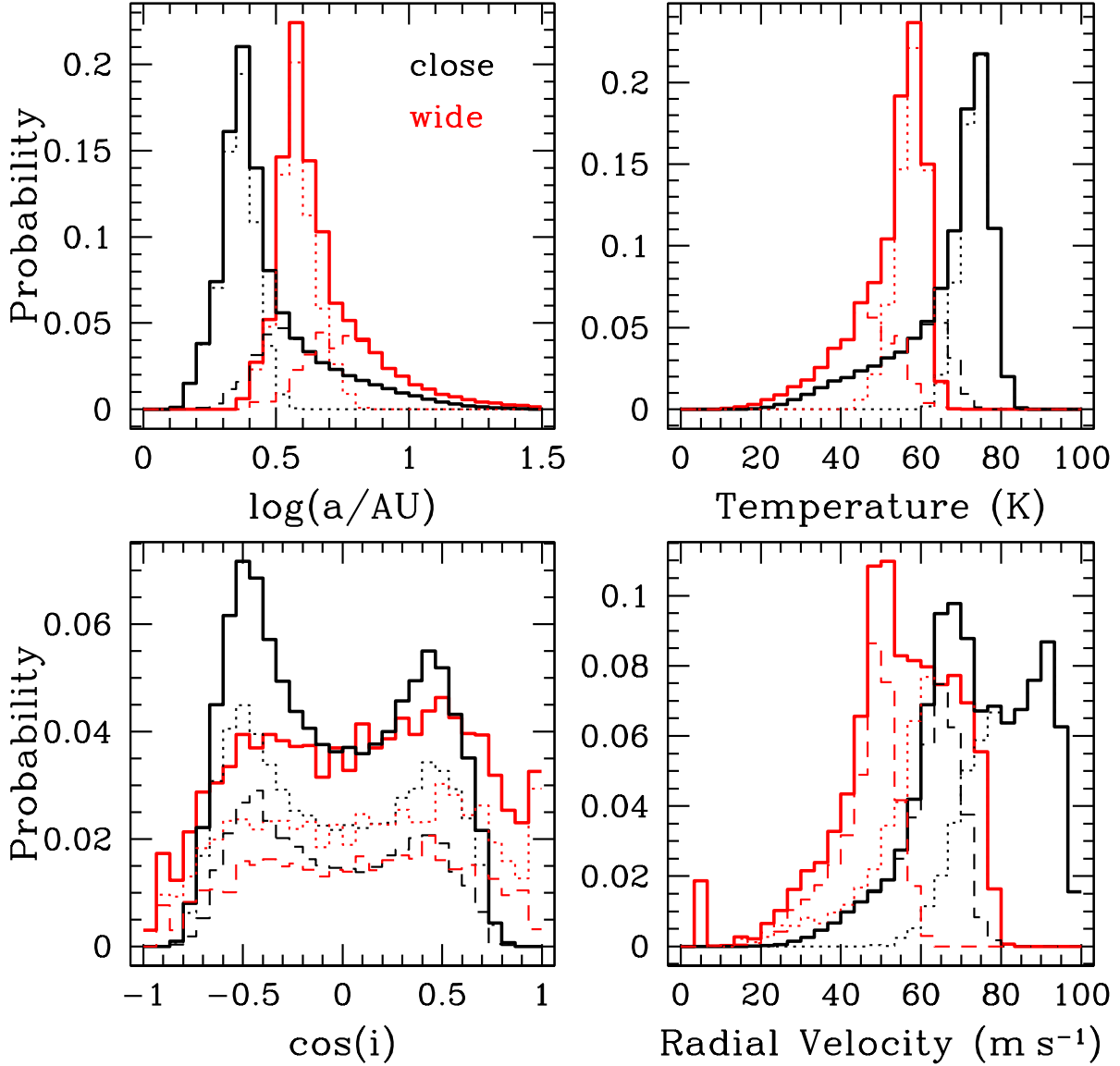}
\caption{Probability distributions of planetary parameters 
(semimajor axis $a$, equilibrium temperature,
cosine of the inclination, and amplitude of radial velocity of
the lens star) from MCMC realizations assuming circular 
orbital motion. Histograms in black and red represent the
close-binary and wide-binary solutions, respectively.
Dotted and dashed histograms represent the two 
degenerate solutions for each MCMC realization discussed in 
Appendix~\ref{sec:appendcirc}.
} 
\label{fig:orbit}
\end{figure}

\paragraph{Circular Planetary Orbits and Planetary Parameters}
\label{sec:circular}
Planetary deviations in microlensing light curves are intrinsically
short, so in most cases, only the instantaneous projected distance
between the planet and the host star can be extracted. As shown in 
\S~\ref{sec:wideclose}, for this event, we tentatively measure the 
instantaneous projected velocity of the planet thanks to the relatively
long $(\sim 4 \,{\rm days})$ duration of the planetary signal. One
cannot solve for the full set of orbital parameters 
just from the instantaneous projected position and velocity.
However, as we show in Appendix~\ref{sec:appendcirc}, we can tentatively
derive orbital parameters by assuming that the planet follows a circular orbit 
around the host star. In Figure~\ref{fig:orbit}, 
we show the probability distributions of the semimajor axis, inclination, 
amplitude of radial velocity, and equilibrium temperature of the planet 
derived from ``MCMC B'' for both wide and close solutions. 
The equilibrium temperature is defined to be 
$T_{\rm eq}\equiv 
(L_{\rm bol}/L_{\rm bol,\odot})^{1/4}(2a/R_\odot)^{-1/2}T_\odot$,
where $L_{\rm bol}$ is the bolometric luminosity of the host, 
$a$ is the planet semimajor axis, and
$L_{\rm bol,\odot}$, $R_\odot$, and $T_\odot$ are the luminosity,
radius, and effective temperature of the Sun, respectively.  This would
give the Earth an equilibrium temperature of $T_{\rm eq} = 285\,$K.
In calculating these probabilities, we assign a flat
(\"{O}pik's Law) prior for the semimajor axis
and assume that the orbits are randomly oriented, that is, 
with a uniform prior on $\cos i$.

\subsection{{Constraints on a Non-Luminous Lens}}
\label{sec:nonlum}

In \S~\ref{sec:hst}, we noted that the blended light must lie within $15\,\mas$
of the source: otherwise the {\it HST} images would appear extended.
We argued that the blended light must be associated with the event 
(either the lens itself or a companion to either the source or lens),
since the chance of such an alignment by a random field star is $<0.07\%$.
In fact, even stronger constraints can be placed on the blend-source
separation using the arguments of \S~\ref{sec:astrometry}.  These are somewhat
more complicated and depend on the blend-source relative parallax, so
we do not consider the general case (which would only be of interest
to further reduce the already very low probability of a random
interloper) but restrict attention to companions of the source and
lens.  We begin with the simpler source-companion case.

\subsubsection{{Blend As Source Companion}
\label{sec:sourcecompanion}}

As we reported in \S~\ref{sec:astrometry}, there were two {\it HST} measurements
of the astrometric offset between the $V$ and $I$ light centroids, dating
from 0.09 and 0.84 years after peak, respectively.  In that section,
we examined the implications of these measurements under the hypothesis
that the blend is the lens.  We therefore ignored the first measurement
because the lens source separation at that epoch is much better constrained
by the microlensing event itself than by the astrometric measurement.
However, as we now examine the hypothesis that the blend is a companion
to the source, both epochs must be considered equally.  Most of the weight
(86\%) comes from the second observation, partly because the astrometric
errors are slightly smaller, but mainly because the blend contributes
about twice the fractional light, which itself reduces the error on the
inferred separation by a factor of 2.  Under this hypothesis, we
find a best-fit source-companion separation of $5\,$mas, with a companion
position angle (north through east) of $280^\circ$.  The (isotropic)
error is $3\,$mas.  Approximating the companion-source
relative motion as rectilinear,
this measurement strictly applies to an epoch 0.73 years
after the event,
but of course the intrinsic source-companion relative motion must be very small compared to
the errors in this measurement.

There would be nothing unusual about such a source-companion projected
separation, roughly $40\pm 25$ AU in physical units.  Indeed,
the local G-star binary distribution function peaks close to this value
\citep{dm92}.

The derived separation is also marginally  
consistent with the companion generating 
a xallarap signal that mimics the parallax signal in our dominant 
interpretation.  The semi-major axis of the orbit would have to be
about 0.8 AU to mimic the 1-year period of the Earth, which corresponds
to a maximum angular separation of about 100 $\mu$as, which is compatible
with the astrometric measurements at the $1.6\,\sigma$ level.

Another potential constraint comes from comparing the color difference with
the magnitude difference of the source and blend.  We find that the
source is about $0.5\pm 0.5$ mag too bright to be on the same main sequence.
However, first, this is only a $1\,\sigma$ difference, which is not
significant. Second, both the sign and magnitude of the difference
are compatible with the source being a slightly evolved turnoff star,
which is consistent with its color.

The only present evidence against the source-companion hypothesis is
that the astrometric offset between $V$ and $I$ {\it HST} images
changes between the two epochs, and that the direction and amplitude
of this change is consistent with other evidence of the proper
motion of the lens.  Since this is
only a $P = 1.7\%$ effect, it cannot be regarded as conclusive.
However, additional {\it HST} observations at a later epoch
 could definitively confirm or rule out this hypothesis.

\subsubsection{{Blend As A Lens Companion}
\label{sec:lenscompanion}}

A similar, but somewhat more complicated line of reasoning essentially
rules out the hypothesis that the blend is a companion to the lens, 
at least if the lens is luminous.
The primary difference is that the event itself places very strong
lower limits on how close a companion can be to the lens.

A companion with separation (in units of $\theta_\e$) $d\gg 1$ induces
a \citet{cr} caustic, which is fully characterized by the gravitational
shear $\gamma = q/d^2$.  We find that the light-curve distortions
induced by this shear would be easily noticed unless $\gamma<0.0035$, 
that is,
\begin{equation}
\gamma = {q_c\over d_c^2} = {q_c\theta_\e^2\over \theta_c^2}< 0.0035,
\label{eqn:thetac}
\end{equation}
where $q_c=M_c/M$ is the ratio of the companion mass to the lens mass and
$d_c=\theta_c/\theta_\e$ is the ratio of the lens-companion 
separation to the Einstein radius.  Equivalently,
\begin{equation}
\theta_c > 19\biggl({q_c\over 1.3}\biggr)^{1/2}\theta_\e.
\label{eqn:thetac2}
\end{equation}
Here, we have normalized $q_c$ to the minimum mass ratio required for
the companion to dominate the light assuming that both are main-sequence
stars.  (We will also consider completely dark lenses below).

We now show that equation (\ref{eqn:thetac2}) is inconsistent with
the astrometric data.
If a lens companion is assumed to generate the blend light, then
essentially the same line of reasoning given in \S~\ref{sec:sourcecompanion}
implies that 0.73 years after the event, this companion lies
5 mas from the source, at position angle $280^\circ$ and with an isotropic
error of 3 mas.  The one wrinkle is that we should now take account of
the relative-parallax term in equation (\ref{eqn:centroid_color}), whereas this
was identically zero (and so was ignored) for the source-companion case.
However, this term is only about $1.8\pi_\rel$ and hence is quite
small compared to the measurement errors for typical $\pi_\rel\la 0.2 {\rm mas}$.
We will therefore ignore this term in the interest of simplicity,
except when we explicitly consider the case of large $\pi_\rel$ further
below.

Of course, the lens itself moves during this interval.  From the
parallax measurement alone (i.e. without attributing the $V/I$
astrometric displacement to lens motion), it is known that the lens
is moving in the same general direction, i.e., with position angle roughly
$210^\circ$.  In assessing the amplitude of this motion we consider
only the constraints from finite-source effects (and ignore the
astrometric displacement).  These constraints yield a hard lower
limit on $\theta_\e$ (from lack of pronounced finite-source
effects) of $\theta_\e>0.6\,$mas, which corresponds to a proper motion
$\mu= 3.1\,\mas\,{\rm yr}^{-1}$.   At this extreme value (and
allowing for $2\,\sigma$ uncertainty in the 
direction of lens motion as well in the measurement of the companion
position), the maximum lens-companion separation is 11.4 mas
(i.e., $19\,\theta_\e$),
which is just ruled out by equation (\ref{eqn:thetac2}). At larger 
$\theta_\e$, the lens-companion scenario is excluded more robustly.
For example, in the limit of large $\theta_\e$, we have
$\theta_c = \mu \times 0.73\,{\rm yr} = \theta_\e(0.73\,{\rm yr}/t_\e)=3.9\theta_\e$,
which is clearly ruled out by equation (\ref{eqn:thetac2}).

Then we note that any scenario involving values of $\pi_\rel$ that are
large enough that they cannot be ignored in this analysis ($\pi_\rel\ga 0.5\,\mas$),
must also have very large $\theta_\e=\pi_\rel/\pi_\e\ga 1\,\mas$, 
a regime in which the lens-companion is easily excluded.

The one major loophole to this argument is that the lens may be a
stellar remnant (white dwarf, neutron star, or black hole), in
which case it could be more massive than the companion despite the
latter's greater luminosity.  

\subsection{{Xallarap Effects and Binary Source}
\label{sec:xallarap}}

\begin{figure}
\plotone{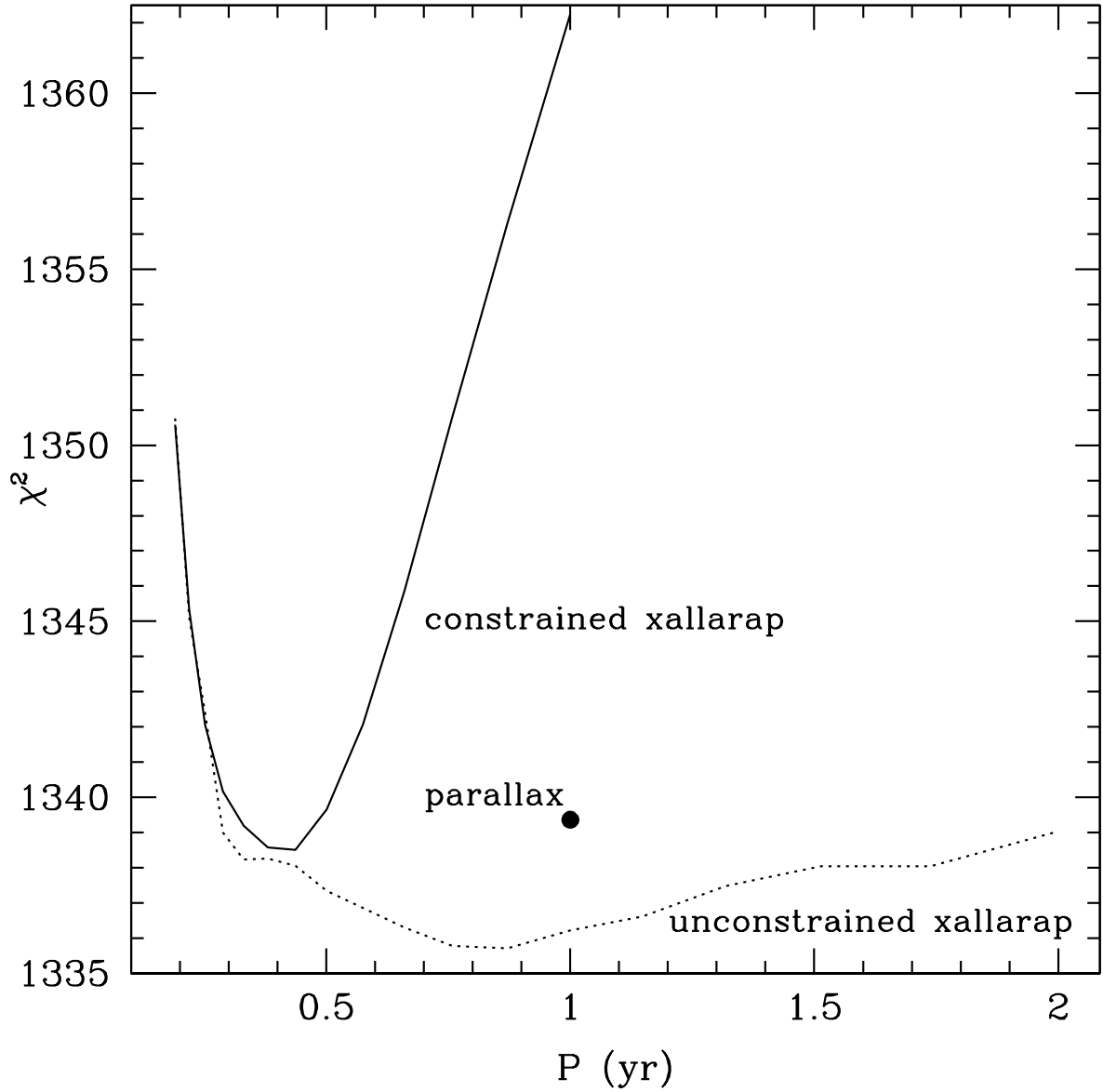}
\caption{$\chi^2$ distributions for best-fit xallarap 
solutions at fixed binary-source orbital periods $P$. 
The solid and dotted
lines represent xallarap fits with and without dynamical constraints described
in \S~\ref{sec:xallarap}. The best-fit parallax solution is shown as a filled
dot at period of 1 year. All of the fits shown in this figure assume 
no planetary orbital motion.}
\label{fig:period}
\end{figure}

\begin{figure}
\plotone{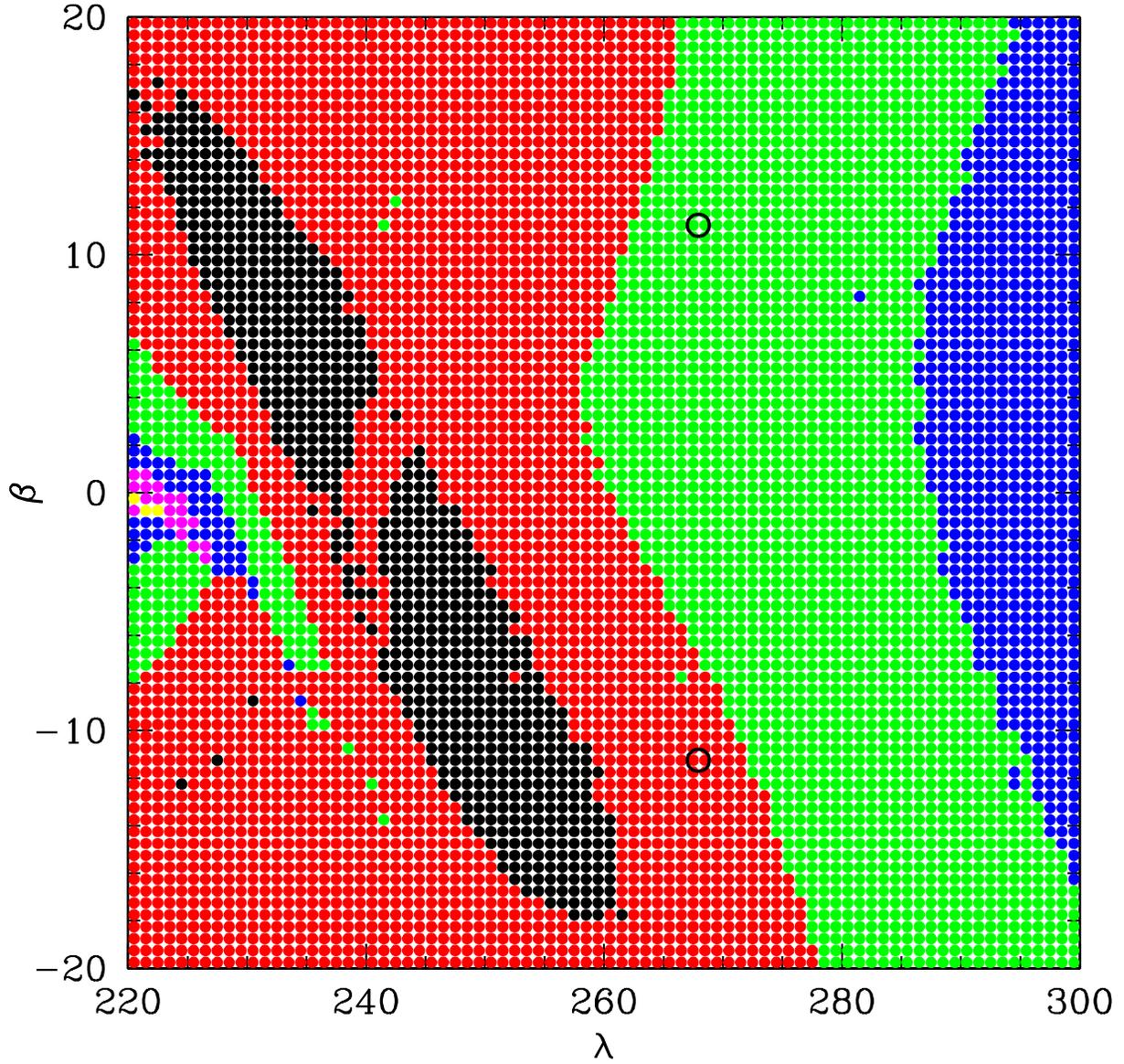}
\caption{
Results of xallarap fits by fixing binary orbital phase 
$\lambda$ and complement of inclination $\beta$ 
at period $P = 1 \, {\rm yr}$ and $u_0>0$. 
The plot is color-coded for 
solutions with $\Delta{\chi^2}$ within 1 (black), 4 (red), 9 (green), 
16 (blue), 25 (magenta), 49 (yellow) of the best fit.
The Earth parameters are indicated by black circles. Because of a
perfect symmetry $(u_0 \rightarrow -u_0$ and $\alpha \rightarrow -\alpha)$, 
the upper black circle represents Earth parameter $(\lambda = 268^{\circ},\, \beta = -11^{\circ})$ for the case $u_0<0$. Comparison of parallax with xallarap
must be made with the better of the two, that is, the lower one.}
\label{fig:lambda_beta}
\end{figure}

Binary source motion can give rise to distortions of the light curve, called
``xallarap'' effects. One can always find a set of xallarap parameters
to perfectly mimic parallax distortions caused by the Earth's motion \citep{smith}.
However, it is a priori unlikely for the binary source to have such parameters, 
so if the parallax signal is real, one would expect the xallarap fits to converge 
to the Earth parameters. For simplicity, we assume that the binary source is in circular orbit. We extensively search the parameter space 
on a grid of 5 xallarap parameters, namely, the period of binary motion 
$P$, the phase $\lambda$ and complement of inclination $\beta$ of the binary orbit, 
which corresponds to the ecliptic longitude and 
latitude in the parallax interpretation of the light curve, as well as
$(\xi_{\rm E,E}, \xi_{\rm E,N})$, which are the counterparts of 
$(\pi_{\rm E,E}, \pi_{\rm E,N})$ of the microlens parallax. We take advantage of the two 
exact degeneracies found by \citet{poindexter05} to reduce the range of 
the parameter search. One exact degeneracy takes $\lambda^{\prime} = \lambda + \pi$ 
and ${\bdv{\chi}_E}^{\prime} = -{\bdv{\chi}_E}$, while all other parameters
remain the same. The other takes $\beta^{\prime} = -\beta$,
${u_0}^{\prime} = -u_0$ and $\xi_{E,N}^{\prime} = -\xi_{E,N}$ (the sign
of $\alpha$ should be changed accordingly as well). 
Therefore we restrict our search to solutions
with positive $u_0$ and with $\pi \leq \lambda \leq 2\pi$. 
In modeling xallarap, planetary orbital motion is neglected.
In Figure~\ref{fig:period}, the $\chi^2$ distribution for best-fit 
xallarap solutions as a function of period is displayed in a dotted line, 
and the xallarap solution with a period of $1$ year has a $\Delta{\chi^2} = 0.5$ 
larger than the best fit at $0.9$ year.
Figure~\ref{fig:lambda_beta} shows that, for the xallarap solutions with 
period of 1 year, the best fit has a $\Delta{\chi^2} = 3.2$ less than the 
best-fit parallax solution (displayed as a black circle point) and its orbital 
parameters are close to 
the ecliptic coordinates of event $(\lambda = 268^{\circ},\, \beta = -11^{\circ})$. 
Therefore, the overall best-fit xallarap solution has  
$\Delta{\chi^2} = 3.7$ smaller than that of the parallax solution 
(whose $\chi^2$ value is
displayed as a filled dot in Fig.~\ref{fig:period}) for  
3 extra degrees of freedom, which gives a probability of $30\%$.
The close proximity between the best-fit xallarap parameters and those of the Earth 
can be regarded as good evidence of the parallax interpretation. The slight 
preference of xallarap
could simply be statistical fluctuation or reflect low-level systematics
in the light curve (commonly found in the analysis by \citealt{poindexter05}).

We also devise another test on the plausibility of xallarap.
In \S~\ref{sec:hst}, we argued that the blend is unlikely to be a random
interloper unrelated to either the source or the lens. If the source were in
a binary, then the blend would naturally be explained as the companion of the
source star. Then from the blend's position on the CMD,
its mass would be $m_c \sim 0.9\,M_\odot$. 
By definition, $\xi_\e$ is the size 
of the source's orbit $a_s$ in the units of $\hat r_\e$ (the Einstein radius 
projected on the source plane),
\begin{equation}
\xi_\e = {a_s\over \hat r_\e} = {{a m_c}\over{(m_c + m_s)\hat r_\e}},
\label{eqn:chie}
\end{equation}
where $a$ is the semimajor axis of the
binary orbit, and $m_s$ and $m_c$ are the masses of the source and
its companion, respectively. Then we apply Kepler's Third Law:
\begin{equation}
\biggl({ P\over \rm yr}\biggr)^2
{m_c^3\over M_\odot(m_c + m_s)^2} =
\biggl({\xi_\e\hat r_\e\over \rm AU}\biggr)^3.
\label{eqn:chie2}
\end{equation}
Once the masses of the source and companion are known, 
the product of $\xi_\e$ and $\hat r_\e$ are determined for a given binary
orbital period $P$. And in the present case, $\hat r_\e / {\rm AU} = \theta_\e 
D_s = \theta_* / \rho D_s = 4.5\times10^{-3}/\rho$. By adopting $m_s = 1 M_\odot$,
$m_c = 0.9 M_\odot$, for each set of $P$ and $\rho$, there is a uniquely
determined $\xi_\e$ from equation~(\ref{eqn:chie2}). We then apply this constraint in
the xallarap fitting for a series of periods. 
The minimum $\chi^2$s for each period from the fittings are shown in 
solid line in Figure~\ref{fig:period}. 
The best-fit solution has $\Delta{\chi^2} \sim 1.0$ less than the best-fit 
parallax solution for two extra degrees of freedom. 
Although as compared to the test described in the previous paragraph, 
 the current test implies a higher probability that the data are explained by parallax 
(rather than xallarap) effects, it still does not rule out xallarap.

\section{{Summary and Future Prospects}
\label{sec:future}}
Our primary interpretation of the OGLE-2005-BLG-071 data assumes that
the light-curve distortions are due to parallax rather than xallarap
and that the blended light is due to the lens itself rather than a
companion to the source. Under these assumptions, the lens is
fairly tightly constrained to be a foreground M dwarf,
with mass $M=0.46\pm 0.04\,M_\odot$ and distance $D_{l} = 3.2\pm 0.4\,$kpc, 
which has thick-disk kinematics ($v_{\rm LSR}\sim 103\,\rm km\,s^{-1}$). 
As we discuss below, future observations might help to constrain its 
metallicity. The microlens modeling suffers from a well-known wide-close 
binary degeneracy. The best-fit wide-binary solutions are slightly 
favored over the close-binary solutions, however, from dynamical 
constraints on planetary orbital motion, the physically allowed solutions
are not distinguishable within $1\,\sigma$. For the wide-binary model, 
we obtain a planet of mass $M_p = 3.8\pm 0.4\,M_{\rm Jupiter}$ at 
projected separation $r_\perp = 3.6\pm 0.2\,$AU. The planet then has an equilibrium temperature of about $T = 55$ K, i.e. similar to Neptune. In the 
degenerate close-binary solutions, the planet is closer to the star and 
so hotter, and the estimates are: $M_p = 3.4\pm 0.4\,M_{\rm Jupiter}$, 
$r_\perp = 2.1\pm 0.1\,$AU and $T \sim 71$ K.

As we have explored in considerable detail, it is possible that
one or both of these assumptions is incorrect.  However, future
astrometric measurements that are made after the lens and source have
had a chance to separate, will largely resolve both ambiguities.
Moreover, such measurements will put much tighter constraints on the
metallicity of the lens (assuming that it proves to be the blended light).

First, the astrometric measurements made 0.84 yr after the event
detected motion suggests that there was still $1.7\%$ chance that the blend 
did not move relative to the source.
A later measurement that detected this motion at higher confidence
would rule out the hypothesis that the blend is a companion to
the source.  We argued in \S~\ref{sec:sourcecompanion} that the blend could not be
a companion to a main-sequence lens.  Therefore, the only possibilities
that would remain are that the lens is the blend, that the lens is
a remnant (e.g., white dwarf), or that the blend is a random interloper
(probability $<10^{-3}$).  As we briefly summarize below, 
a future astrometric measurement could strongly constrain
the remnant-lens hypothesis as well.

Of course, it is also possible that future astrometry will reveal that
the blend does not moving with respect to the source, in which case
the blend would be a companion to the source.  Thus, either way, these
measurements would largely resolve the nature of the blended light.

Second, by identifying the nature of blend, these measurements will
largely, but not entirely, resolve the issue of parallax vs.\ xallarap.
If the blend proves not to be associated with the source, then
any xallarap-inducing companion would have to be considerably less
luminous, and so (unless it were a neutron star) less massive than
the $m_c=0.9\,M_\odot$ that we assumed in evaluating
equation (\ref{eqn:chie2}).  Moreover, stronger constraints on $\hat r_\e$
(rhs of eq.~[\ref{eqn:chie2}]) would be available from the astrometric
measurements.  Hence, the xallarap option would be either excluded
or very strongly constrained by this test.

On the other hand, if the blend were confirmed to be a source companion,
then essentially all higher order constraints on the nature of the lens
would disappear.  The parallax ``measurement'' would then very plausibly be
explained by xallarap, while the ``extra information'' about $\theta_\e$
that is presently assumed to come from the blend proper-motion measurement
would likewise evaporate.

These considerations strongly argue for making a future high-precision
astrometric measurement.   Recall that in the {\it HST} measurements reported
in \S~\ref{sec:hst}, the source and blend were not separately resolved:
the relative motion was inferred from the offset between the
$V$ and $I$ centroids, which are displaced because the source and
blend have different colors. Due to its well-controlled
PSF, {\it HST} is capable of detecting the broadening of the PSF even if
the separation of the lens and source is a fraction of the FWHM. Assuming 
that the proper motion is $\mu_{\rm geo} \sim 4.4\,\masyr$, and based 
on our simulations in \S~\ref{sec:hst}, such broadening would be
confidently detectable about $5$ years after the event 
(see also \citealt{bennett07} for analytic PSF broadening estimates). 
Ten years after the event, the net 
displacement would be $\sim 40\,{\rm mas}$. This compares to a 
diffraction limited FWHM of 40 mas for $H$ band on a ground-based 10m 
telescope and would therefore enable full resolution. The $I-H$ color of
the source is extremely well determined ($0.01\,{\rm mag}$) from simultaneous 
$I$ and $H$ data taken during the event from the CTIO/SMARTS 1.3m in Chile.
Hence, the flux allocation of the partially or fully resolved blend and source stars
would be known. The direct detection of a partially or fully resolved lens will provide
precise photometric and astrometry measurements (see \citealt{direct} for one such
example), which will enable much tighter constraints on the mass, 
distance and projected velocity of the lens. 
It also opens up the possibility of determining the metallicity
of the host star by taking into account both non-photometric and photometric 
constraints. If, as indicated by the projected velocity measurement, it is 
a thick-disk star, then it will be one of the few such stars found to harbor a planet \citep{haywood}.

As remarked above, a definitive detection of the blend's proper motion would
still leave open the possibility that it was a companion to the
lens, and not the lens itself.  In this case, the lens would have to be
a remnant.  Without going into detail, the astrometric measurement
would simultaneously improve the blend color measurement as well
as giving a proper-motion estimate (albeit with large errors because the
blend-source offset at the peak of the event would then not be known).
It could then be asked whether the parallax, proper-motion, and photometric
data could be consistently explained by any combination of
remnant lens and main-sequence companion.  This analysis would depend
critically on the values of the measurements, so we do not explore
it further here.  We simply note that this scenario could also be
strongly constrained by future astrometry.

\section{{Discussion}
\label{sec:discussion}}
With the measurements presented here, and the precision with which
these measurements allow us to determine the properties of the planet
OGLE-2005-BLG-071Lb and its host, it is now possible to place this
system in the context of similar planetary systems discovered by
radial velocity (RV) surveys.  Of course, the kind of information that
can be inferred about the planetary systems discovered via RV differs
somewhat from that presented here.  For example, for planets discovered
via RV, it is generally only possible to infer a lower limit to the
planet mass, unless the planets happen to transit or produce a
detectable astrometric signal. {\it Mutatis mutandis}, for planets discovered via
microlensing, it is generally only possible to measure the projected
separation at the time of the event, even in the case for which the
microlensing mass degeneracy is broken as it is here (although see
\citealt{ob06109}).

With these caveats in mind, we can compare the properties of
OGLE-2005-BLG-071Lb and its host star with similar RV systems.  It is
interesting to note that the fractional uncertainties in the host mass and
distance of OGLE-2005-BLG-071Lb are comparable to those of some of the
systems listed in Table \ref{tab:planets}.  

OGLE-2005-BLG-071Lb is one of only eight Jovian-mass  
($0.2 M_{\rm Jupiter} < M_p < 13 M_{\rm Jupiter}$) planets that have
been detected orbiting M dwarf hosts (i.e., $M_*<0.55~M_\odot$)
\citep{marcy98,marcy01,delfosse98,butler06,johnson07b,bailey}.  Table \ref{tab:planets}
summarizes the planetary and host-star properties of the known M
dwarf/Jovian-mass planetary systems.
OGLE-2005-BLG-071Lb is likely the most massive known planet orbiting an
M dwarf.

As suggested by the small number of systems listed in Table
\ref{tab:planets}, and shown quantitatively by several recent studies,
the frequency of relatively short-period $P \la 2000~{\rm days}$,
Jupiter-mass companions to M dwarfs appears to be $\sim 3-5$ times
lower than such companions to FGK dwarfs
\citep{butler06,endl06,johnson07b,cumming08}. This paucity, which has
been shown to be statistically significant, is expected in the
core-accretion model of planet formation, which generally predicts
that Jovian companions to M dwarfs should be rare, since for lower
mass stars, the dynamical time at the sites of planet formation is
longer, whereas the amount of raw material available for planet
formation is smaller (\citealt{laughlin04,ida05,kennedy08}, but see
\citealt{kornet06}).  Thus, these planets typically do not reach
sufficient mass to accrete a massive gaseous envelope over the lifetime
of the disk.  Consequently, such models also predict that in the outer
regions of their planetary systems, lower mass stars should host a
much larger population of `failed Jupiters,' cores of mass $\la
10~M_\oplus$ \citep{laughlin04,ida05}.  Such a population was indeed
identified based on two microlensing planet discoveries
\citep{ob05390,ob05169}.

Our detection of a $\sim 4\,M_{\rm Jupiter}$ companion to an M dwarf
may therefore present a difficulty for the
core-accretion scenario.  While we do not have a constraint on the
metallicity of the host, the fact that it is likely a member of the
thick disk suggests that its metallicity may be subsolar.  If so, this
would pose an additional difficulty for the core-accretion scenario,
which also predicts that massive planets should be rarer around
metal-poor stars \citep{ida04}, as has been demonstrated observationally
\citep{santos04,fischer05}.  This might imply that a different mechanism
is responsible for planet formation in the OGLE-2005-BLG-071L system, such
as the gravitational instability mechanism \citep{boss02,boss06}.

One way to escape these potential difficulties is if the host lens is
actually a stellar remnant, such as white dwarf.  The progenitors of
remnants are generally more massive stars, which are both predicted
\citep{ida05,kennedy08} and observed \citep{johnson07a,johnson07b} to
have a higher incidence of massive planets.  As we discussed above,
future astrometric measurements could constrain both the
low-metallicity and remnant-lens hypotheses.  These measurements are
therefore critical.

Although it is difficult to draw robust conclusions from a single
system, there are now four published detections of Jovian-mass
planetary companions with microlensing (\citealt{ob03235,ob06109}), 
and several additional such planets have been detected that are currently being 
analyzed. It is therefore reasonable to expect several detections per year 
\citep{gould09}, and thus that it will soon be possible to use microlensing to 
constrain the frequency of massive planetary companions.  These constraints are
complementary to those from RV, since the microlensing detection
method is less biased with respect to host star mass \citep{gould00a},
and furthermore probes a different region of parameter space, namely
cool planets beyond the snow line with equilibrium temperatures
similar to the giant planets in our solar system (see,
e.g.~\citealt{gould07} and \citealt{gould09}).

\acknowledgments
We thank Marc Pinsonneault and Deokkeun An for providing us their 
unpublished isochrones. S.D. wishes to thank David Will of Ohio State astronomy 
department for setting up and maintaining the Condor system, which greatly 
facilitates the computations for this work. S.D. is grateful to
Ondrej Pejcha and David Heyrovsky  for interesting discussions on limb-darkening.
Based on observations with the NASA/ESA Hubble Space 
Telescope obtained at the Space Telescope Science Institute, which 
is operated by the  Association of Universities for Research in Astronomy, 
Incorporated, under NASA contract NAS5-26555. Support for this work was 
provided by NASA through grant HST-GO-10707.01-A from STScI. 
S.D. and A.G. were supported in part by grant AST 042758 from the NSF. 
S.D., A.G., D.D. and R.P. acknowledge support by NASA grant NNG04GL51G.
AG thanks IAP,CNRS for its support.
Support for OGLE project was provided by Polish MNiSW grant N20303032/4275.
B.G.P. was supported by the grant (KRF-2006-311-C00072) from
Korea Research Foundation. HC was supported by the Science
Research Center from Korea Science and Engineering Foundation.
The MOA project is supported by Ministry of Education, Culture, Sports, 
Science and Technology (MEXT) of Japan, Grant-in-Aid for Specially Promoted 
Research No. 14002006. 
JPB, PF, AC, CC, SB, JBM acknowledge the financial support of ANR HOLMES.
KHC's work performed under the auspices of the U.S. Department of Energy by
Lawrence Livermore National Laboratory under Contract DE-AC52-07NA27344.
This work was supported in part by an allocation of computing time 
from the Ohio Supercomputer Center.

\appendix
\section{Extracting Orbital Parameters for Circular Planetary Orbit}
\label{sec:appendcirc}
OGLE-2005-BLG-071 is the first planetary microlensing event for
which the effects of planetary orbital motion in the light curve have been fully
analyzed.
The distortions of the light curve due to the orbital motion are modeled
by $\omega$ and $\dot b$ as discussed in \S~\ref{sec:orbit}. In
addition, the lens mass $M$ and distance $D_l$ are determined,
so we can directly convert the microlens light-curve parameters that are 
normalized to the Einstein radius to physical parameters. 
In this section, we show that under the assumption of a circular planetary orbit, 
the planetary orbital parameters 
can be deduced from the light-curve parameters. Let $r_\perp = D_l \theta_\e d$ be 
the projected star-planet separation and let $r_\perp \bdv{\gamma}$ be the 
instantaneous planet velocity in the plane of the sky, i.e.
$r_\perp \gamma_\perp = r_\perp \omega$ is the velocity perpendicular to this axis and
$r_\perp \gamma_\parallel = r_\perp \dot d/d$ is the velocity parallel to this 
axis. Let $a$ be the semi-major axis and 
define the $\bdv{{\hat{\imath}}}$, $\bdv{\hat{\jmath}}$, $\bdv{\hat{k}}$ directions 
as the instantaneous star-planet-axis on the sky plane, the direction
 into the sky, and 
 $\bdv{\hat{k}} = \bdv{{\hat{\imath}}} \times \bdv{\hat{\jmath}}$.
Then the instantaneous velocity of the planet is
\begin{equation}
{\bf v} = \sqrt{GM\over a} [\cos\theta {\bdv{\hat {k}}} + 
\sin\theta (\cos\phi \bdv{\hat{\imath}} - \sin\phi \bdv{\hat{\jmath}})],
\label{eqn:velocity}
\end{equation}
where $\phi$ is the angle between star-planet-observer (i.e., 
$r_\perp = a\sin\phi$) and
$\theta$ is the angle of the velocity relative to the $\bdv{\hat{k}}$ direction
on the plane that is perpendicular to the planet-star-axis. We thus obtain
\begin{equation}
\gamma_\perp = \sqrt{GM\over a^3}{\cos\theta\over \sin\phi},\qquad
\gamma_\parallel = \sqrt{GM\over a^3}{\sin\theta\cot\phi}.
\label{eqn:gamma}
\end{equation}
To facilitate the derivation, we define
\begin{equation}
A\equiv {\gamma_\parallel\over\gamma_\perp} = -\tan\theta \cos\phi, \qquad
B\equiv {r_\perp^3\gamma_\perp^2\over GM} = \cos^2\theta\sin\phi,
\label{eqn:aandb}
\end{equation}
which yield as an equation for $\sin\phi$:
\begin{equation}
B = F(\sin\phi); \qquad
F(x) = {x(1-{x}^2)\over A^2 + 1 - {x}^2}.
\label{eqn:sinphi}
\end{equation}
Note that $F'(\sin\phi) = 0$ when 
$\sin^2\phi_* = (3/2)A^2 + 1 - |A|\sqrt{(9/4)A^2 + 2}$
. So equation~(\ref{eqn:sinphi}) has two degenerate solutions when 
$B < F(\sin \phi_*)$ and has no solutions when $B > F(\sin \phi_*)$.
Subsequently, one obtains,
\begin{equation}
a = {r_\perp\over{\sin \phi}}, \qquad
\cos i = -\sin\phi \cos\theta, \qquad
K = \sqrt{GM \over a} q \sin{i},
\label{eqn:solutions}
\end{equation}
where $i$ is the inclination and $K$ is the amplitude of radial velocity.

The Jacobian matrix used to transform 
from $P(r_\perp,\gamma_\perp,\gamma_\parallel)$ to $P(a,\phi,\theta)$
is given below,

\begin{eqnarray}
{\p(r_\perp,\gamma_\perp,\gamma_\parallel)\over
\p(a,\phi,\theta)} 
\nonumber &=&
{GM\over a^3}\Bigg|\matrix{\sin\phi & a\cos\phi & 0\cr
-{3\over 2a}{\cos\theta\over\sin\phi} &
-{\cos\theta\cos\phi\over\sin^2\phi} & - {\sin\theta\over\sin\phi} \cr
-{3\over 2a}{\sin\theta\cot\phi} &
-{\sin\theta\over\sin^2\phi} & \cos\theta\cot\phi}\Bigg| 
\nonumber\\ && 
={GM\over a^3}\cot^2\phi\biggl({1\over 2}-\sin^2\theta\tan^2\phi\biggr).
\label{eqn:jacobi}
\end{eqnarray}

{ Then for an arbitrary function $H(a)$,
\begin{equation}
{\p(r_\perp,\gamma_\perp,\gamma_\parallel)\over
\p(H(a),\cos\phi,\theta)} =
{\p(r_\perp,\gamma_\perp,\gamma_\parallel)\over
\p(a,\phi,\theta)}\times {1\over \sin\phi H'(a)},
\label{eqn:ha}
\end{equation}
which, for the special case of a flat distribution, $H(a) = \ln a$, yields,
\begin{equation}
{\p(r_\perp,\gamma_\perp,\gamma_\parallel)\over
\p(ln(a),\cos\phi,\theta)}
=  {GM\over r_\perp^2}{\cos^2\phi\over\sin\phi}.
\biggl({1\over 2}-\sin^2\theta\tan^2\phi\biggr)
\label{eqn:jacobi2}
\end{equation}}


\headsep = 105pt
\begin{sidewaystable}
\caption{\label{tab:models}\sc Light Curve Parameter Estimations From Markov chain Monte Carlo Simulations.}
\vskip 1em
MCMC A
\vskip 1em
\begin{tabular}{@{\extracolsep{0pt}}cccccccccccccccc} 
\hline \hline
Model & {$t_0$} & {$u_0$} & {$t_{\rm E}$} & {$d$} & {$q$} & {$\alpha$} & {$\rho$} & {$\pi_{{\rm E},N}$} & {$\pi_{{\rm E},E}$}
& {$\omega$} & {$\dot d/d$} & {$I_{{\rm s}}$} & {$I_{{\rm b}}$} & {$V_{{\rm s}}$} & {$V_{{\rm b}}$} \\
$\chi^2$    & (HJD') & & (day) & &$\times10^3$ & (deg) &$\times 10^4$ & & & (yr$^{-1}$)& (yr$^{-1}$) &(mag) &(mag) &(mag) &(mag)\\ \hline \hline
   Wide$+$&3480.7024& 0.0282&71.1&1.306&7.5&273.63&3.9&-0.30&-0.26&-1.328&-0.256&19.51&21.29&20.85&23.11   \\
   1345.0&$^{+0.0058}_{-0.0054}$&$^{+0.0008}_{-0.0009}$&$^{+2.3}_{-2.4}$&$^{+0.002}_{-0.004}$&${\pm 0.2}$ &$^{+0.16}_{-0.15}$&$^{+1.8}_{-2.7}$&$^{+0.24}_{-0.28}$&${\pm 0.05}$ &$^{+0.274}_{-0.165}$&$^{+0.134}_{-0.129}$&$^{+0.04}_{-0.03}$&$^{+0.22}_{-0.17}$&${\pm 0.04}$ &$^{+0.43}_{-0.26}$\\ \hline
      Wide$-$&3480.7028&-0.0283&70.6&1.307&7.5& 86.21&3.9&-0.34&-0.26& 1.117&-0.277&19.51&21.30&20.85&23.11   \\
      1345.3&$^{+0.0054}_{-0.0056}$&$^{+0.0010}_{-0.0008}$&${\pm 2.2}$ &$^{+0.003}_{-0.004}$&${\pm 0.3}$ &$^{+0.13}_{-0.16}$&$^{+1.8}_{-2.6}$&$^{+0.30}_{-0.23}$&${\pm 0.05}$ &$^{+0.130}_{-0.293}$&$^{+0.152}_{-0.113}$&${\pm 0.04}$ &${\pm 0.19}$ &${\pm 0.03}$ &$^{+0.43}_{-0.26}$\\ \hline
        Close$+$&3480.6789& 0.0239&70.1&0.763&6.9&274.27&3.1&-0.36&-0.27& 0.301& 0.502&19.52&21.28&20.85&23.13   \\
	1345.8&$^{+0.0055}_{-0.0043}$&$^{+0.0009}_{-0.0007}$&$^{+2.1}_{-2.4}$&$^{+0.004}_{-0.006}$&${\pm 0.3}$ &$^{+0.25}_{-0.36}$&$^{+1.7}_{-2.5}$&$^{+0.24}_{-0.27}$&${\pm 0.05}$ &$^{+0.486}_{-0.788}$&$^{+0.148}_{-0.101}$&${\pm 0.03}$ &$^{+0.21}_{-0.16}$&$^{+0.04}_{-0.03}$&$^{+0.40}_{-0.28}$\\ \hline
	  Close$-$&3480.6799&-0.0241&69.2&0.762&6.9& 85.53&2.7&-0.33&-0.26&-0.405& 0.528&19.52&21.30&20.85&23.17   \\
	  1345.2&$^{+0.0042}_{-0.0051}$&$^{+0.0009}_{-0.0007}$&$^{+2.3}_{-1.9}$&$^{+0.004}_{-0.006}$&${\pm 0.3}$ &$^{+0.39}_{-0.26}$&${\pm 2.2}$ &$^{+0.28}_{-0.26}$&${\pm 0.05}$ &$^{+0.696}_{-0.622}$&$^{+0.127}_{-0.118}$&${\pm 0.03}$ &${\pm 0.18}$ &${\pm 0.04}$ &$^{+0.35}_{-0.32}$\\ \hline
\end{tabular}

\vskip 1em
MCMC B
\vskip 1em
\begin{tabular}{@{\extracolsep{0pt}}cccccccccccccccc} 
\hline \hline
Model & {$t_0$} & {$u_0$} & {$t_{\rm E}$} & {$d$} & {$q$} & {$\alpha$} & {$\rho$} & {$\pi_{{\rm E},N}$} & {$\pi_{{\rm E},E}$}
& {$\omega$} & {$\dot d/d$} & {$I_{{\rm s}}$} & {$I_{{\rm b}}$} & {$V_{{\rm s}}$} & {$V_{{\rm b}}$} \\
$\chi^2$    & (HJD') & & (day) & &$\times10^3$ & (deg) &$\times 10^4$ & & & (yr$^{-1}$)& (yr$^{-1}$) &(mag) &(mag) &(mag) &(mag)\\ \hline \hline
   Wide$+$&3480.7015& 0.0287&69.3&1.305&7.7&273.67&6.1&-0.02&-0.22&-1.242&-0.283&19.49&21.40&20.82&23.91   \\
   1353.4&$^{+0.0050}_{-0.0059}$&${\pm 0.0007}$ &$^{+1.6}_{-1.7}$&$^{+0.003}_{-0.005}$&${\pm 0.2}$ &$^{+0.17}_{-0.11}$&${\pm 0.4}$ &${\pm 0.12}$ &${\pm 0.03}$ &$^{+0.321}_{-0.125}$&${\pm 0.129}$ &${\pm 0.03}$ &${\pm 0.19}$ &${\pm 0.03}$ &$^{+0.24}_{-0.20}$\\ \hline
      Wide$-$&3480.7012&-0.0287&69.2&1.305&7.7& 86.29&6.0& 0.02&-0.21& 1.193&-0.293&19.49&21.40&20.82&23.97   \\
      1353.3&$^{+0.0052}_{-0.0060}$&${\pm 0.0007}$ &$^{+1.7}_{-1.8}$&$^{+0.002}_{-0.005}$&${\pm 0.2}$ &$^{+0.12}_{-0.15}$&$^{+0.5}_{-0.3}$&$^{+0.10}_{-0.13}$&${\pm 0.03}$ &$^{+0.127}_{-0.342}$&$^{+0.131}_{-0.121}$&${\pm 0.02}$ &${\pm 0.19}$ &${\pm 0.03}$ &$^{+0.19}_{-0.24}$\\ \hline
        Close$+$&3480.6792& 0.0245&68.3&0.763&7.0&274.38&6.0&-0.01&-0.22& 0.415& 0.569&19.49&21.35&20.83&23.88   \\
	1355.5&$^{+0.0041}_{-0.0051}$&$^{+0.0005}_{-0.0006}$&${\pm 1.6}$ &$^{+0.003}_{-0.006}$&$^{+0.3}_{-0.2}$&$^{+0.23}_{-0.39}$&${\pm 0.4}$ &$^{+0.12}_{-0.15}$&${\pm 0.02}$ &$^{+0.503}_{-0.744}$&$^{+0.112}_{-0.130}$&${\pm 0.03}$ &$^{+0.20}_{-0.16}$&${\pm 0.02}$ &$^{+0.24}_{-0.18}$\\ \hline
	  Close$-$&3480.6793&-0.0245&68.2&0.762&7.1& 85.63&6.0& 0.04&-0.22&-0.179& 0.561&19.50&21.36&20.83&23.90   \\
	  1355.5&$^{+0.0042}_{-0.0051}$&${\pm 0.0006}$ &$^{+1.8}_{-1.5}$&$^{+0.004}_{-0.006}$&${\pm 0.3}$ &$^{+0.46}_{-0.24}$&${\pm 0.4}$ &$^{+0.09}_{-0.16}$&${\pm 0.03}$ &$^{+0.703}_{-0.722}$&$^{+0.126}_{-0.112}$&${\pm 0.02}$ &${\pm 0.18}$ &${\pm 0.02}$ &${\pm 0.21}$ \\ \hline
\end{tabular}
\end{sidewaystable}

\begin{table}
\caption{\label{tab:physical} \sc Derived Physical Parameters}
\vskip 1em
\begin{tabular}{@{\extracolsep{0pt}}ccccccccc}
\hline
\hline
Model & {$M$} & {$\pi_{\rm rel}$} & {$D_l$} & {$\mu_{\rm N}$} & {$\mu_{\rm E}$} & {$\theta_{\rm E}$} & {$M_p$} & {$r_{\perp}$} \\
$\chi^2$    & $M_\odot$ & mas & kpc &mas$\,{\rm yr^{-1}}$ &mas$\,{\rm yr^{-1}}$ & mas &$M_{\rm Jupiter}$ &AU \\ \hline
\hline
   Wide$+$&0.46&0.19&3.2&-0.4&-4.3&0.84&3.8&3.6   \\
1353.4&${\pm 0.04}$ &${\pm 0.04}$ &${\pm 0.4}$ &$^{+2.7}_{-3.1}$&${\pm 0.3}$ &$^{+0.06}_{-0.04}$&$^{+0.3}_{-0.4}$&${\pm 0.2}$ \\ \hline
   Wide$-$&0.46&0.19&3.2& 0.3&-4.3&0.85&3.8&3.6   \\
1353.3&${\pm 0.04}$ &$^{+0.04}_{-0.03}$&${\pm 0.4}$ &$^{+2.3}_{-3.6}$&$^{+0.3}_{-0.2}$&${\pm 0.05}$ &$^{+0.3}_{-0.4}$&${\pm 0.2}$ \\ \hline
  Close$+$&0.46&0.19&3.1&-2.6&-4.4&0.86&3.4&2.1   \\
1355.5&${\pm 0.04}$ &$^{+0.04}_{-0.03}$&${\pm 0.4}$ &$^{+4.8}_{-1.1}$&${\pm 0.3}$ &${\pm 0.05}$ &$^{+0.3}_{-0.4}$&${\pm 0.1}$ \\ \hline
  Close$-$&0.46&0.20&3.1&-0.2&-4.4&0.87&3.4&2.1   \\
1355.5&${\pm 0.04}$ &${\pm 0.04}$ &${\pm 0.3}$ &$^{+3.6}_{-3.4}$&${\pm 0.3}$ &${\pm 0.04}$ &${\pm 0.3}$ &${\pm 0.1}$ \\ \hline
\end{tabular}
\end{table}

\clearpage
\begin{table}
\caption{\label{tab:planets}\sc Jovian-mass Companions to M Dwarfs ($M_* < 0.55~M_\odot$)}
\vskip 1em
\begin{tabular}{@{\extracolsep{0pt}}cccccccc}
\hline
\hline
Name & {$M_*$} & {Metallicity} & {Dist.} & {$M_p$} & {$P$} & {$a$} & {Ref.}\\
&{($M_\odot$)} & & (pc) & {($M_{\rm Jup}$)} & {(days)} & {(AU)} &\\
\hline
\hline
GJ 876c  & $0.32$  & $-0.12$ & $4.660$ & $ 0.6-$     & $30.340$ & $0.13030$ & 1,2,3\\
   & $\pm 0.03$  & $ \pm 0.12$ & $\pm0.004$ & $0.8$ & $\pm 0.013$ & & \\
\hline
GJ 876b  & --               & --               & --            & $1.9 - $        & $60.940$ & $0.20783$ & --\\
         &                &                &            &  2.5                 & $\pm 0.013$       &           & \\
\hline
GJ 849b  & $0.49 $  & $0.16 $  &  $8.8$                & $0.82/\sin i$ & $1890$ & $2.35$ & 4\\
         & $     \pm 0.05$  & $     \pm 0.2$  & $\pm 0.2$           &               & $    \pm 130$ &        &  \\
\hline
GJ 317b   & $0.24 $  & $-0.23 $ & $9.2 $ & $1.2/\sin i$ & $692.9$ & $0.95$ & 5\\
          & $     \pm 0.04$  & $\pm 0.2$ & $\pm 1.7$ &  & $\pm 4$ &  & \\
\hline
GJ 832b   & $0.45 $  & $\sim -0.7 $ & $\sim 4.93 $ & $0.64/\sin i$ & $3416$ & $3.4$ & 6\\
          & $     \pm 0.05$  & /-0.3 &  &  & $\pm 131$ & $\pm0.4$ & \\
\hline
OGLE-2006 & $0.50$ & ? & $1490 $ & $0.71 $ & $1830
$ & $2.3 $ & 7 \\
-BLG-109Lb & $\pm 0.05$ & ? & $ \pm 130$ & $ \pm 0.08$ & $
\pm 370$ & $ \pm 0.2$ &  \\
\hline
OGLE-2006 & -- & -- & -- & $0.27 $ & $5100$ & $4.6 $
& -- \\
-BLG-109Lc &  &  &  & $\pm 0.03$ & $\pm 730$ & $ \pm 0.5$
&  \\
\hline
OGLE-2005 & $0.46$ & Subsolar?\tablenotemark{a} & $3300$ & $3.8 $\tablenotemark{b} & -- & $3.6 $\tablenotemark{b,c} & This\\
-BLG-071Lb & $\pm 0.04$ &  & $ \pm 300$ & $\pm 0.4$ &  & $\pm 0.2$ & Paper\\
\hline
\tablenotetext{a}{While the metallicity of the OGLE-2005-BLG-071Lb host star is not directly constrained
by our data, its kinematics indicate it is likely a member of the metal-poor thick disk.}
\tablenotetext{b}{We give the planet mass and projected separation for the wide solution, which
is favored by $\Delta \chi^2 = 2.1$.  The second, close solution
has $M_p=3.4 \pm 0.3~M_{\rm Jupiter}$ and $r_\perp=2.1 \pm 0.1~{\rm AU}$.}
\tablenotetext{c}{We give the  the projected separation between the host and planet at the time
of event, which is the orbital parameter most directly constrained by our observations.
However, assuming a circular orbit, we infer that the semi-major
axis is likely only $\sim 10 - 20\%$ larger $[a {\rm(wide)} \sim 4.1{\rm AU}, a {\rm(close)} \sim 2.5 {\rm AU}]$.}
\tablerefs{
(1)\citealt{rivera05};  (2)\citealt{bean06}; (3)\citealt{benedict02}; (4)\citealt{butler06}; (5) \citealt{johnson07b} (6) \citealt{bailey} (7) \citealt{ob06109}
}
\end{tabular}
\end{table}
\end{document}